\documentclass[preprint,floatfix,showpacs,pra,preprintnumbers,amsmath,amssymb,superscriptaddress]{revtex4}
\usepackage{epsfig}
\usepackage{color}
\usepackage{amsmath}
\usepackage{amssymb}
\usepackage{citesort}
\usepackage{psfrag}
\usepackage{pstricks}
\usepackage{longtable}
\usepackage{dcolumn}
\usepackage{bm}

\newcommand{\ket}[1]{\mathinner{|{#1}\rangle}}

\newcommand{\braket}[2]{\langle #1|#2\rangle}

\def\b#1{\textcolor{black}{#1}}

\begin{document}

\title{On the construction of model Hamiltonians for adiabatic quantum
  computation and its application to finding low energy conformations of lattice
  protein models}

\author{Alejandro Perdomo}

\affiliation{Department of Chemistry and Chemical Biology, Harvard University,
12 Oxford Street, 02138, Cambridge, MA}

\author{Colin Truncik}

\affiliation{D-Wave Systems, Inc., 4401 Still Creek Drive, Suite 100, Burnaby, BC V5C 6G9, Canada}

\author{Ivan Tubert-Brohman}

\affiliation{Department of Chemistry and Chemical Biology, Harvard University,
12 Oxford Street, 02138, Cambridge, MA}

\author{Geordie Rose}

\affiliation{D-Wave Systems, Inc., 4401 Still Creek Drive, Suite 100, Burnaby, BC V5C 6G9, Canada}

\author{Al\'an Aspuru-Guzik}
\affiliation{Department of Chemistry and Chemical Biology, Harvard University,
12 Oxford Street, 02138, Cambridge, MA}
\email{aspuru@chemistry.harvard.edu}

\begin{abstract}
  In this report, we explore the use of a quantum optimization algorithm for
  obtaining low energy conformations of protein models. We discuss mappings
  between protein models and optimization variables, which are in turn mapped to
  a system of coupled quantum bits. General strategies are given for constructing
  Hamiltonians to be used to solve optimization problems of
  physical/chemical/biological interest via quantum computation by adiabatic
  evolution. As an example, we implement the Hamiltonian
  corresponding to the Hydrophobic-Polar (HP) model for protein folding.
  Furthermore, we present an approach to reduce the resulting Hamiltonian to
  two-body terms gearing towards an experimental realization.
\end{abstract}



\maketitle

\section{Introduction}

Finding the ensemble of low-energy conformations of a peptide given its primary
sequence is a fundamental problem of computational biology, commonly known
as the protein folding problem
\cite{Chan1993,Plotkin2002,Shakhnovich2006,Mirny2001,Dill1995,Gruebele1999,Creighton1992}.
The native fold conformation is usually assumed to correspond to the global minimum of the protein's free energy (according to the so-called thermodynamic
hypothesis \cite{Epstein1963}), although some exceptions have been proposed
\cite{Baker1994zapo34,Lazaridisa2003zapo33}. Thus, the protein folding
problem can be described as a global optimization problem. Algorithms for
quantum computers have been developed for many applications such as factoring
\cite{Shor1997} and the calculation of molecular energies
\cite{Aspuru-Guzik2005}. In this report, we investigate the approach of using an
adiabatic quantum computer for folding a highly simplified protein model.

\begin{figure}[h]
 \begin{center}
 \includegraphics[scale=0.4]{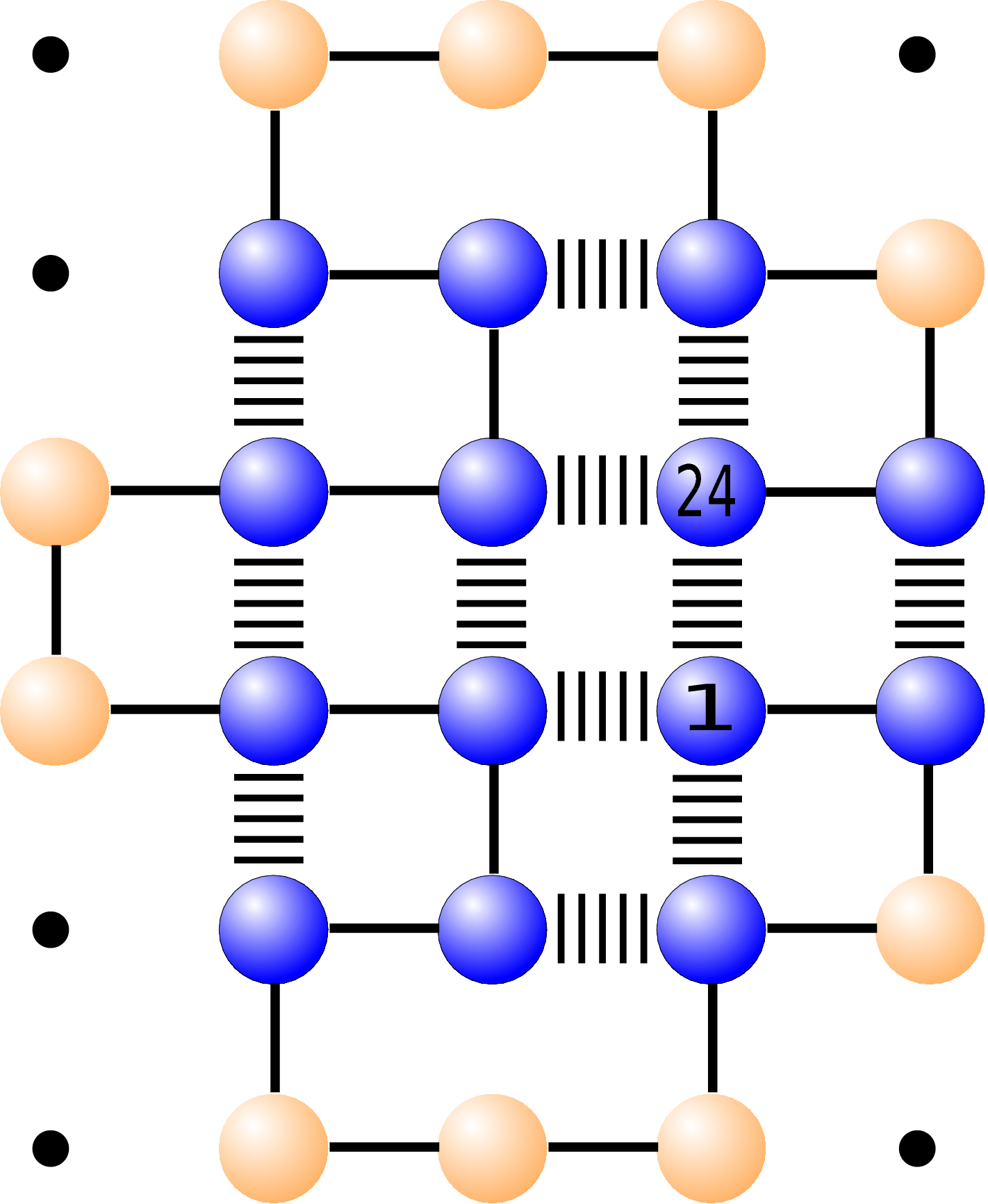}
\end{center}
\caption{\label{fig:hp-model}(Color online) The lattice protein hydrophobic-polar (HP) model,
  showing the global energy minimum conformation for a sequence of 24 amino
  acids, HHPHPPPHHHHPPHHHHPPPHPHH $(E=-12)$. Blue (dark grey) beads represent
  hydrophobic residues (H) and orange (light grey) beads represent polar residues (P).
  The model consists of a self-avoiding chain with favorable ($E=-1$) energetic
  interactions among hydrophobic residues in contact. Contact between nearest neighbors in the primary sequence are unavoidable, and their contribution is not added to the calculated energy. Black
  dots represent lattice sites. Dotted lines represent favorable energetic interactions, solid lines represent the self-avoiding chain.}
\end{figure}

The HP (H: hydrophobic, P: polar) lattice model \cite{LAU1989} is one of the
simplest protein models implemented. Still its accuracy in predicting some of the folding behaviour of real
proteins has made it a useful benchmark for testing optimization
algorithms such as simulated annealing \cite{Steinhofel2007}, genetic algorithms
\cite{Cox2004,Cox2006,Custodio2004,Unger1993,Song2005}, and ant colony
optimization \cite{Shmygelska2005}. Other heuristic methods such as hydrophobic
core threading \cite{Backofen2003}, chain growth \cite{Beutler1996,Hsu2003},
contact interactions \cite{Toma1996}, and hydrophobic zippers \cite{Yue1995}
have also been considered. The HP model has also been useful for a qualitative
investigation of the nature of the folding process and the interactions between
proteins. The HP model depicted in Fig.~\ref{fig:hp-model} is defined by
three assumptions: 1) There are only two kinds of amino acids or
residues, hydrophobic (H) and polar (P); 2) residues are placed on a
grid (typically a square grid for the 2D model and a cubic grid for the 3D
model); 3) the only interaction among amino acids is the favorable contact
between two H residues that are not adjacent in the sequence. The energy of this interaction is defined as -1 in arbitrary units, representing a hydrophobic effect which tends to fold the protein in a way that aggregates the H residues in a predominantly hydrophobic core, and leaves the P residues at the surface of the protein. The search for the
native conformation of the protein is represented by a self-avoiding walk on the
grid.

An important property of the model is that the number of possible conformations
is roughly proportional to $2.7^N$ \cite{LAU1989}, where $N$ is the length of
the polypeptide chain. Proofs of the NP-completeness of both the 2D and 3D
HP models have been given \cite{Papadimitriou1998,Berger1998}.
Due to this exponential growth, global optimality proofs become impractical
when $N$ reaches approximately 50 residues. For
longer sequences, heuristics and stochastic algorithms have been employed for
$N$ up to 136 for the 3D HP model \cite{Toma1996}.

This report is structured as follows. Sec.~\ref{sec:implementation} presents the general quantum algorithm and the terms of the Hamiltonian necessary to obtain the folded structure of the protein, and
describes how to map the problem to arrays of coupled
quantum bits \cite{KAMINSKY2004,Harris2007}. Sec.~\ref{sec:contruction-h-terms}
explains the construction of the core component of the algorithm, the
Hamiltonian that encodes the lowest energy conformation of the protein. In Sec.~\ref{sec:n4d2} we solve in detail the four amino acid
sequence HPPH in a two-dimensional grid. In Sections~\ref{sec:2local} and
~\ref{sec:resources} we discuss the resources necessary to carry out the reduction
from a general $k$-body Hamiltonian to a two-body Hamiltonian, as a function of
the size of the protein.

\section{An adiabatic quantum algorithm for the HP model}\label{sec:implementation}

We begin this section by describing the mapping of a sequence of $N$ amino acids into binary variables, which will in turn be mapped to spin variables in the quantum mechanical version of the algorithm.

\b{\subsection{Mapping amino acids onto a lattice}\label{subsec:mapping}}

\begin{figure}[h]
\begin{center}
 \includegraphics[scale=0.4]{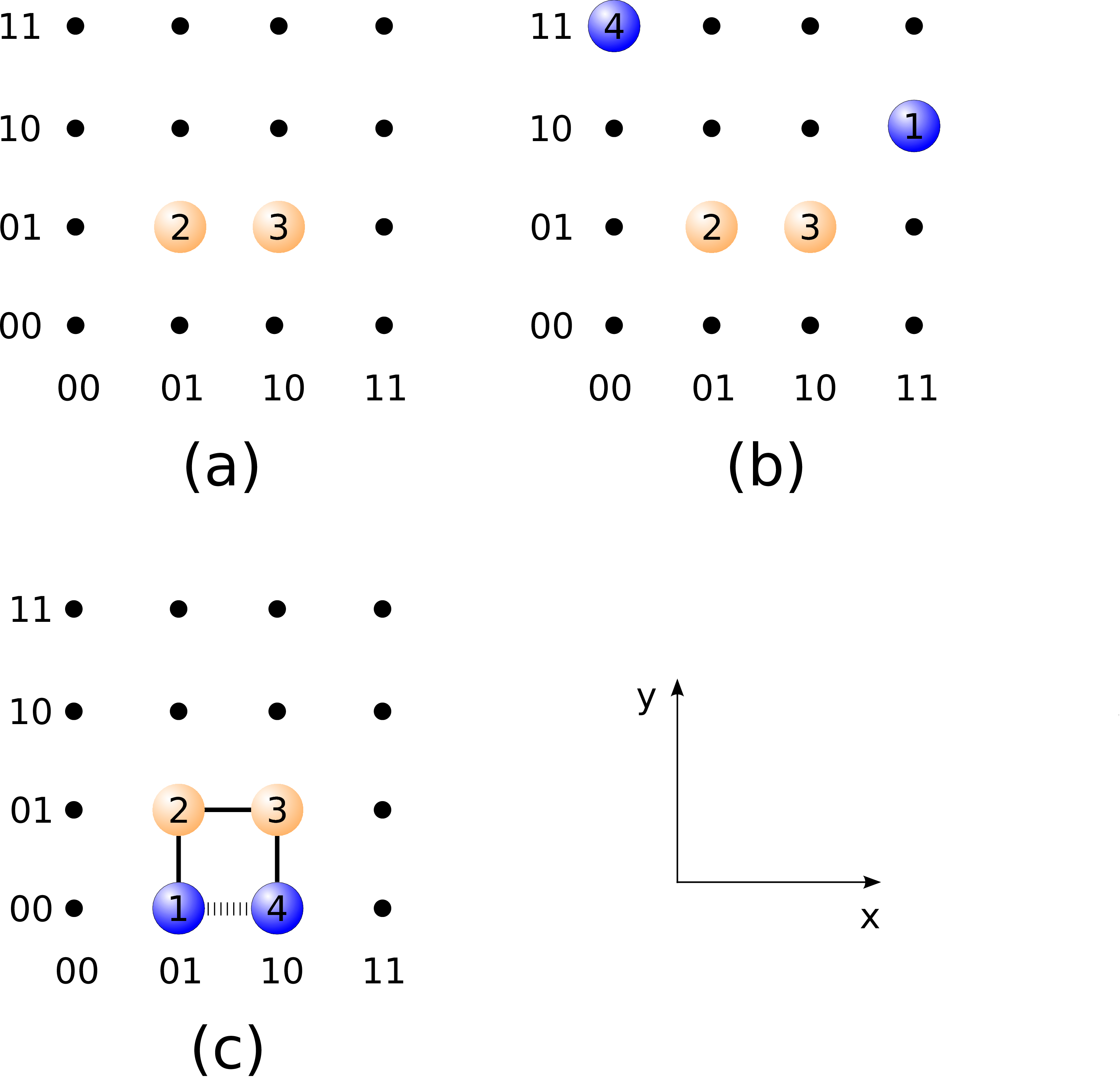}
\end{center}
\caption{\label{fig:2Dgrid}(Color online) Grid-labeling conventions for a sequence of 4 amino acids, HPPH. {\bf (a)} Amino acids 2 and 3 are fixed in the center of the grid to eliminate translational degeneracy. {\bf (b)} One of the possible invalid configurations that might arise in the search and that would need to be discarded by the optimization algorithm. {\bf (c)} Lowest-energy conformation for this example. The dotted line between amino acids 1 and 4 represents the hydrophobic interaction favored by the HP model. The configurations to 
  optimize assume the form $q = q_{16} q_{15} q_{14} q_{13}\,\,0110\,\,0101\,\,q_{4} q_{3} q_{2} q_{1}$, where the set of variables
  $q_{16} q_{15} q_{14} q_{13}$ and $q_{4} q_{3} q_{2} q_{1}$ determine the position of amino acids 4 and 1, respectively. For the particular case in (b), $q = 1100\,\,0110\,\,0101\,\,1011$. }
\end{figure}

\b{The mapping of the coordinates of a sequence of $N$ amino acids to a given grid
of size $N \times N$ is developed as follows. We assume, without loss of
generality, that the number of amino acids is a power of 2. A binary representation for the labels of the grid
requires $\log_2 N$ binary variables to specify the position of an amino acid in
each dimension, as shown in Fig.~\ref{fig:2Dgrid}. The position of each of $N$ amino acids in a $D$-dimensional lattice may thus be encoded by a bit string $q$ composed of exactly $D N \log_2 N$ binary variables $q_i$. For example, for $N=4$, $D=2$,
the length of the bit string $q$ is 16 and therefore the number of
configurations that can be explored is $2^{16}$. Let $q$ denote a particular
configuration of the protein in the grid, written in the form}
\begin{equation}\label{q}
  q = \underbrace{q_{16} q_{15}}_{y_4}\,\,\underbrace{q_{14} q_{13}}_{x_4}\,\,\underbrace{q_{12} q_{11}}_{y_3}\,\,\underbrace{q_{10} q_{9}}_{x_3}\,\,\underbrace{q_{8} q_{7}}_{y_2}\,\,\underbrace{q_{6} q_{5}}_{x_2}\,\,\underbrace{q_{4} q_{3}}_{y_1}\,\,\underbrace{q_{2} q_{1}}_{x_1},
\end{equation}
where $x_i$ and $y_i$ are the $x$ and $y$ coordinate of the $i$-th amino acid.
Fig.~\ref{fig:2Dgrid} shows an example of the coordinate mapping given a
specific sequence of residues or amino acids.

In the quantum version of the problem, these configurations span a Hilbert space of dimension $2^{16}$. The state
vectors can be written as
\begin{equation}
\ket{q} \equiv \ket{q_{16}} \ket{q_{15}} \cdots  \ket{q_{2}} \ket{q_{1}}.
\end{equation}
We wish to implement a Hamiltonian which encodes the ground state
of the protein on a spin-$1/2$ quantum computer \cite{DiVincenzo2000}, or, in
particular onto an Ising-like Hamiltonian with a transverse magnetic field
\cite{Bravyi2007} (see Sec.~\ref{sec:experiment}). To do so, we realize the $16$-qubit Hilbert space
as a system of $16$ spin-$1/2$ particles, with $\ket{q_{i} = 0}$
mapped to the spin state $\ket{\sigma_{i}^{z} = +1}$ and $\ket{q_{i} = 1}$
mapped to $\ket{\sigma_{i}^{z} = -1}$, with these spin states as the computational basis. In other words, the quantum version of
the configuration states is related to spin variables through the
transformation
\begin{equation}\label{eq:qz-operator}
 \hat q_{i} \equiv \frac{1}{2} (I - \hat{\sigma}_{i}^{z}),
\end{equation}
with $I = \bigl( \begin{smallmatrix} 1 & 0 \\ 0 & 1 \end{smallmatrix} \bigr)$ and $\sigma^{z} = \bigl( \begin{smallmatrix} 1 & 0 \\ 0 & -1 \end{smallmatrix} \bigr)$, the identity operator and the $\sigma^{z}$ Pauli matrix represented in the computational basis, respectively.

In Sec.~\ref{sec:contruction-h-terms} we will derive an energy function in terms of the $N D \log_2 N$ binary variables used to describe all of the possible configurations for the $N$ amino acids in a $D$-dimensional lattice. This energy function is constructed so that its minimum will yield the lowest-energy conformations of the protein. Eq.~\ref{eq:qz-operator} provides the rule for the mapping of this energy function to a quantum Hamiltonian. Each $q_i$ in the energy function will be replaced by an operator $\hat q_{i}$. The operator $\hat q_{i}$ is to be understood as a short hand notation for a quantum operator acting on the $i$-th qubit of the $N D \log_2 N$ multipartite Hilbert space, $\mathcal{H}_{ND log_2 N} \otimes \mathcal{H}_{ND log_2 N - 1} \otimes \cdots \otimes \mathcal{H}_i \otimes \cdots \otimes \mathcal{H}_1$. The explicit form of $\hat q_i$ is given by $I \otimes I \otimes \cdots \otimes \hat q_i \otimes \cdots \otimes I$. Notice that the operator $\hat q_i$ as defined in Eq.~\ref{eq:qz-operator} has been placed in the $i$-th position, and the identity operator acts on the rest of the Hilbert space. Products of the form $q_i q_j$ will be replaced by a quantum operator $\hat q_i \hat q_j$, which is a shorthand notation for the operators $\hat q_i$ and $\hat q_j$ acting on the $i$-th and the $j$-th qubits, respectively. As an illustrative example, consider an energy function dependent on four binary variables,
\[E(q_1, q_2, q_3, q_4)= 1 - q_1 q_2 + q_1 q_3 + q_2 q_3 q_4, \]
which will be mapped to a Hamiltonian acting on a four qubit Hilbert space, $\mathcal{H}_4 \otimes \mathcal{H}_3 \otimes \mathcal{H}_2 \otimes \mathcal{H}_1$. In the instance of this particular energy function the Hamiltonian will assume the form
\begin{align}
 \hat H &= I \otimes I \otimes I \otimes I - I \otimes I \otimes \hat q \otimes \hat q + I \otimes \hat q \otimes I \otimes \hat q + \hat q \otimes \hat q \otimes \hat q \otimes I \nonumber\\ &\equiv I -\hat q_1 \hat q_2 + \hat q_1 \hat q_3 + \hat q_2 \hat q_3 \hat q_4.
\end{align}
 
Following this mapping, transformation of any energy function to the quantum Hamiltonian is straightforward.

In order to eliminate redundancy due to translational symmetry, we fixed the two
middle amino acids in a central position (see Fig.~\ref{fig:2Dgrid}). This reduces the number of binary variables in the bit string from sixteen to eight. The variables corresponding to amino acids 1 and 4: $q_{4} q_{3} q_{2} q_{1}$ and $q_{16}
q_{15} q_{14} q_{13}$, respectively, become the variables of interest, and the
variables $q_{8} q_{7} q_{6} q_{5}$ and $q_{12} q_{11} q_{10} q_{9}$
corresponding to amino acids 2 and 3, become constant throughout the optimization process. In general, the $(N/2)^{th}$ amino acid is assigned to the $(N/2)^{th}$ grid point in all $D$ dimensions. The $(N/2+1)^{th}$ amino acid is fixed to the $(N/2+1)^{th}$ grid point in
the $x$ direction and to the $(N/2)^{th}$ grid point in all other $D-1$
dimensions. As shown in Fig.~\ref{fig:2Dgrid}, the final configuration we will
try to optimize for the case of four amino acids takes the form $\ket{q} =
\ket{q_{16} q_{15} q_{14} q_{13}} \ket{0110} \ket{0101} \ket{q_{4} q_{3} q_{2}
  q_{1}}$.

\subsection{Adiabatic Quantum Computation}\label{sec:experiment}

The goal of an adiabatic quantum algorithm is to
transform an initial state into a final state which encodes the answer to the
problem. A quantum state $\ket{\psi(t)}$ in the $2^n$-dimensional Hilbert space for $n$ qubits, evolves in time according to the Schr\"{o}dinger equation
\begin{equation}
 i\hbar \frac{d}{dt}\ket{\psi(t)} = \hat H(t) \ket{\psi(t)},
\end{equation}
where $\hat H(t)$ is the time-dependent Hamiltonian operator. The design of the algorithm takes advantage of the quantum adiabatic theorem
\cite{Messiah}, which is satisfied whenever $\hat H(t)$ varies slowly throughout
the time of propagation $t \in [0, \tau]$. Let $\ket{\psi_{g} (t)}$ be the
instantaneous ground state of $\hat H(t)$. If we construct $\hat H(t)$ such that the
ground state of $\hat H(0)$, denoted as $\ket{\psi_{g} (0)}$, is easy to prepare, the
adiabatic theorem states that the time propagation of the quantum state will
remain very close to $\ket{\psi_{g} (t)}$ for all $t \in [0, \tau]$. One way to choose
$\hat H(0)$ is to construct it in such a way that $\ket{\psi_{g} (0)}$ is a
uniform superposition of all possible configurations of the system, i.e.
\begin{equation} \label{psi0}
 \ket{\psi_{g} (0)} = \frac{1}{\sqrt{2^{n}}} \sum_{q_{i} \in \{0,1\}} \ket{q_{n}} \ket{q_{n-1}} \cdots \ket{q_2} \ket{q_1}
\end{equation}
summing over all $2^{n}$ vectors of the computational basis. Notice that an initial Hamiltonian of the form
\begin{equation}
 \hat H(0) = \sum_{i=1}^{n} \hat q_x^{i} = \sum_{i=1}^{n} \frac{1}{2}(I-\hat{\sigma}^{x}_{i})
\end{equation}
would have as a non-degenerate ground state the vector $\ket{\psi_{g} (0)}$ defined in Eq.~\ref{psi0}.

Similarly to the operator $\hat q$ from Eq.~\ref{eq:qz-operator}, we define
\begin{equation}\label{eq:qx-operator}
 \hat q_x^{i} \equiv \frac{1}{2}(I-\hat{\sigma}^{x}_{i}),\end{equation}
with $I = \bigl( \begin{smallmatrix} 1 & 0 \\ 0 & 1 \end{smallmatrix} \bigr)$ and $\sigma^{x} = \bigl( \begin{smallmatrix} 0 & 1 \\ 1 & 0 \end{smallmatrix} \bigr)$, the identity operator and the $\sigma^{x}$-Pauli matrix represented in the computational basis, respectively.

For example, for the case of four qubits, $n=4$, $\hat H(0)$ is given by,
\begin{align}
 \hat H(0) &= \sum_{i=1}^{4} \hat q_x^{i} = \hat q_x^{1}+ \hat q_x^{2}+ \hat q_x^{3} + \hat q_x^{4} \\
& =  I \otimes I \otimes I \otimes \hat q_x + I \otimes I \otimes \hat q_x \otimes I + I \otimes \hat q_x \otimes I \otimes I + \hat q_x \otimes I \otimes I \otimes I.
\end{align}
To find the lowest energy conformation of the protein, one defines a
Hamiltonian, $\hat H_{protein}$, whose ground state encodes the solution. Adiabatic evolution begins with $\hat H(0)$ and $\ket{\psi_{g} (0)}$, and ends in
$\hat H_{protein}= \hat H(\tau)$. If the adiabatic evolution is slow enough, the state obtained at
time $t=\tau$ is $\ket{\psi_{g} (\tau)}$, the ground state
of $\hat H(\tau)= \hat H_{protein}$. The details about the construction of $\hat H_{protein}$
will be provided in Sec.~\ref{sec:contruction-h-terms}. A possible adiabatic
evolution path can be constructed by the linear sweep of a parameter $t \in
[0,\tau]$,
\begin{equation}\label{h(t)}
 \hat H(t) = (1-t/\tau) \hat H(0) + (t/\tau) \hat H_{protein}.
\end{equation}
Even though Eq.~\ref{h(t)} connects $\hat H(0)$ and $\hat H_{protein}$, determining the
optimum value of $\tau$ is an important and non-trivial problem in itself. In
principle, the adiabatic theorem states that over sufficient adiabatic time
$\tau$, the state $\ket{\psi(\tau)}$ will converge to the solution to the problem
$\ket{\psi_{g} (\tau)}$. The magnitude of $\tau$ dictates the ultimate
usefulness of the quantum algorithm proposed in this work. Farhi \textit{et al.}
\cite{Farhi2000,Farhi2001} showed promising numerical results for random
instances of the Exact Cover computational problem.

Notice that the parameter $\tau$ determines the rate at which $\hat H(t)$ varies.
Following the notation from Farhi {\it et al} \cite{Farhi2000}, consider $\hat H(t)=
\tilde{H} (t/ \tau) = \tilde{H}(s)$, with instantaneous values of
$\tilde{H}(s)$ defined by
\begin{equation}
 \tilde{H}(s)\ket{l;s}=E_{l}(s)\ket{l;s}
\end{equation}
with
\begin{equation}
 E_{0}(s) \le E_{1}(s) \le \cdots \le E_{N-1}(s)
\end{equation}
where $N$ is the dimension of the Hilbert space. According to the adiabatic theorem, if the gap between the two lowest levels, $E_{1}(s) - E_{0}(s)$, is greater than zero for all $0 \le s \le 1$, and taking
\begin{equation}
 \tau \gg \frac{\varepsilon}{g_{min}^{2}}
\end{equation}
with the minimum gap, $g_{min}^{2}$, defined by
\begin{equation}
 g_{min}= \min_{0 \le s \le 1} (E_{1}(s) - E_{0}(s)),
\end{equation}
and $\varepsilon$ given by
\begin{equation}
 \varepsilon = \max_{0 \le s \le 1} \vert \braket{l=1;s}{\frac{d\tilde{H}}{ds} \vert l=0;s} \mid,
\end{equation}
then we can make
\begin{equation}
 \vert \braket{l=0;s=1}{\psi(\tau)} \vert
\end{equation}
arbitrarily close to 1. In other words, the existence of a nonzero gap guarantees that $\ket{\psi (t)}$ remains very close to the ground state of $\hat H(t)$ for all $0 \le t \le \tau$, if $\tau$ is sufficiently large.

In the following sections, we derive the expression for an energy function which is mapped to $\hat H_{protein}$ using the procedure explained in Sec~\ref{subsec:mapping}. The final expression for $\hat H_{protein}$ corresponds to an array of coupled qubits.  We use $H$ to denote both the Hamiltonians and the energy functions given that the mapping is straightforward as explained at the end of Sec.~\ref{subsec:mapping}.

\section{Construction of the lattice protein Hamiltonian for adiabatic quantum computation}\label{sec:contruction-h-terms}
Our goal in this section is to find an algebraic expression for an energy function in which the ground state represents the lowest energy conformation of a protein. Ideally,
this energy function should contain the least possible number of terms. In order to
optimize the computational resources, we desire terms with low
locality, defined as the number of products of $q_{i}$'s that appear
in a certain term (e.g., a term of the form $h=q_1 q_3 q_4 q_6$ is 4-local).

\subsection{Small computer science digression}\label{sec:comp_alg}
Encoding positions of the amino acids in the grid in terms of Boolean
variables makes it very convenient to use tools from computer science and basic
Boolean algebra \cite{Rosen1999}. In this section, we will review these tools
before using them to contruct arbitrary Hamiltonians
that encode the spectrum of statistical mechanical models. We begin with some
simple relations that are useful in the derivation of the Hamiltonian terms.

\begin{figure}[h]
\begin{center}
 \includegraphics[scale=0.5]{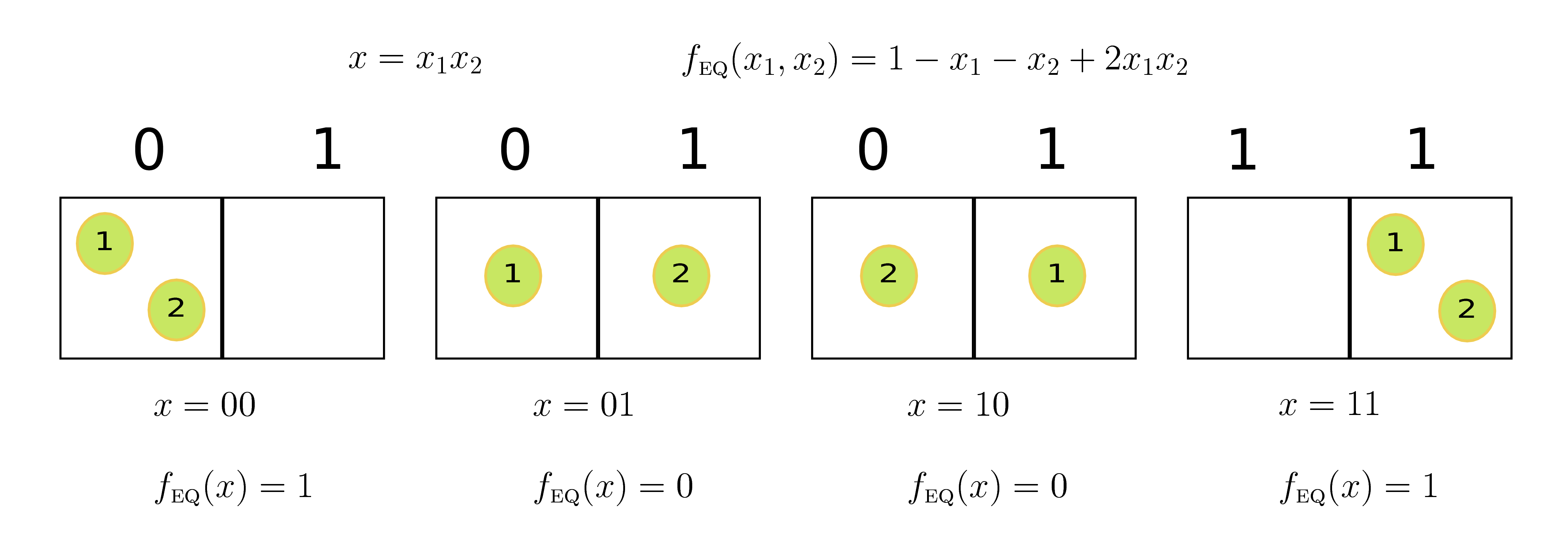}
\end{center}
\caption{\label{fig:equality} (Color online) Illustrative example of one of the uses of the
  {\sc xnor} Boolean function in our scheme for the construction of
  Hamiltonians. Consider two particles 1 and 2 that are restricted to
  occupy either position 0 or 1 in the dimension shown, and let $x_1$ and $x_2$ encode the position
  particle 1 and particle 2 respectively. The Boolean function $f_{EQ}$ can be
  interpreted as an onsite repulsion Hamiltonian which penalizes configurations
  where $x_1 = x_2$. The possible configurations are encoded in the bit string
  $x=x_1 x_2$.}
\end{figure}

Consider two Boolean variables $x$ and $y$. Expressions for the operations {\sc and}, {\sc
  or}, {\sc not} can be written algebraically as:
\begin{align*}
&f_{\textrm{{\sc and}}}(x,y) =  xy  &\textrm{{\sc and} operation} \: (x \wedge y)\\
&f_{\textrm{{\sc or}}}(x,y)=x+y-xy  &\textrm{{\sc or} operation} \: (x \vee y)\\
&f_{\textrm{{\sc not}}}(x) = 1-x  &\textrm{{\sc not} operation} \: (\neg x)
\end{align*}
An additional useful Boolean operator for the construction of Hamiltonian terms is {\sc xnor}. The output of the {\sc xnor} function is 0 unless all its arguments have the same value. The two-input version
{\sc xnor} operation is also known as {\sl logical equality}, here denoted as {\sc EQ},
\begin{equation*}
f_{\textrm{{\sc eq}}}(x,y)= 1-x-y+2xy \qquad\quad \textrm{{\sc xnor} operation}(x \: \textrm{{\sc eq}} \: y)
\end{equation*}
The {\sc xnor } operator can be used to construct a very useful term for
statistical mechanics Hamiltonians, an on-site repulsion penalty (described in Sec.~\ref{subsec:hamiltoniansprotein} and illustrated in Fig.~\ref{fig:equality}).

\subsection{Hamiltonian terms for protein folding: the HP model}\label{subsec:hamiltoniansprotein}
Most of the configurations represented by the bit strings $q$ of Eq.~\ref{q}
are invalid protein states. We seek a Hamiltonian that
energetically favors valid configurations of the HP model by eliminating
configurations in which more than one amino acid occupy the same grid point, and discarding
configurations that violate the primary sequence of amino acids. This
Hamiltonian can be written as
\begin{equation}\label{h-protein}
H_{protein} = H_{onsite} + H_{psc} + H_{pairwise},
 \end{equation}
where $H_{onsite}$ is an onsite repulsion term for amino acids occupying the same grid point, $H_{psc}$ is a primary sequence
constraint term,  and $H_{pairwise}$ is a pairwise interaction term that
represents favorable hydrophobic interactions between adjacent hydrophobic amino acids.

Each protein configuration can be described by a string of $ND\log_2 N$ bits,
where $D$ is the number of dimensions and $N$ is the number of amino acids. Without loss of generality, $N$ is here contrained to be a power of two. Below, we describe each term in Eq.~\ref{h-protein}.

\subsubsection{Onsite term, $H_{onsite}$}
\label{sec:onsite}
The first term in Eq.~\ref{h-protein}, $H_{onsite}$, prevents two or more amino
acids from occupying the same grid point. For a given protein, at least one position variable must differ
between each pair of amino acids for $H_{onsite}$ to evaluate to zero. As an
illustrative example, a simple one-dimensional two-site Hamiltonian is shown in
Fig.~\ref{fig:equality} using the {\sc xnor} operation described in Sec.~\ref{sec:comp_alg}.

The general term for $D$ dimensions and $N$ amino acids is
\begin{equation}\label{h-onsite}
H_{onsite}(N,D) = \lambda_0 \sum_{i=1}^{N-1} \sum_{j=i+1}^{N} H_{onsite}^{ij}(N,D)
 \end{equation}
with
\begin{eqnarray}\label{h-onsite-ij_app}
H_{onsite}^{ij}(N,D) = &&\prod_{k=1}^{D} \prod_{r=1}^{\log_2 N} \Bigl(1-q_{f(i,k)+r}-q_{f(j,k)+r}\nonumber\\ &&+2\,q_{f(i,k)+r}\,q_{f(j,k)+r}\Bigr)
 \end{eqnarray}
and
\begin{equation}\label{f_app}
f(i,k)=D(i-1)\log_2 N +(k-1)\log_2 N.
\end{equation}
The terms enclosed by the parentheses in Eq. \ref{h-onsite-ij_app} are {\sc xnor} functions. The double product of these terms tests that all of these conditions are considered simultaneously by using {\sc and} relations. If all the binary variables describing the coordinates of the $i$-th and $j$-th amino acids are equal, then the series of products of {\sc xnor} functions is evaluated  to +1. In this case, the energy penalty $\lambda_0$ with $\lambda_0 > 0$ is enforced. There will be no energy penalty, however, if even one of the binary variables for the $i$-th and $j$-th amino acids is different.

The function $f(i,k)$ is a pointer to the bit substring describing the coordinates of a particular amino acid. The index $i$ points to the $i$-th amino acid and the index $k$ points to the first bit variable of the $k$-th spatial coordinate. Here, $k=1$ corresponds to the $x$ coordinate, $k=2$ to the $y$ coordinate, and $k=3$ to the $z$ coordinate. For example, consider the case with $N=4$ and $D=2$. If we are interested in referring to the first binary variable describing the $y$ coordinate ($k = 2$), for the third amino acid ($i=3$), a direct substitution in Eq.~\ref{f_app} would yield $f(3,2) = 10$, which is indeed the variable we are interested in according to the convention established in Eq.~\ref{q}. 



\subsubsection{Primary structure constraint, $H_{psc}$}
\label{sec:psc}
The term $H_{psc}$ in Eq.~\ref{h-protein} evaluates to zero when two amino acids
$P$ and $Q$ that are consecutive sequence-wise must be nearest
neighbors on the lattice. Nearest-neighbors are defined as those points with a rectilinear ($L_1$) distance of
$d_{PQ}=1$ between them. We define a distance function that gives the base 10 distance squared between any two amino
acids $P$ and $Q$ on the lattice,
\begin{equation}\label{eq:dist_ap}
d^{2}_{PQ}(N, D) = \sum_{k=1}^{D} \Bigl(\sum_{r=1}^{\log_2 N} 2^{r-1}(q_{f(P,k)+r}-q_{f(Q,k)+r}) \Bigr)^2
\end{equation}
with $f(i,k)$ defined as in Eq.~\ref{f_app}.

A simple way of defining $H_{psc}$ is
\begin{equation}\label{h`-psc_ap}
H^{\prime}_{psc}(N,D) = \lambda_1 \sum_{m=1}^{N-1} (1-d^{2}_{m,m+1})^2
\end{equation}
Or, preferably,
\begin{equation}\label{h-psc_ap}
H_{psc}(N,D) = \lambda_1 \Bigl[-(N-1) + \sum_{m=1}^{N-1} d^{2}_{m,m+1}\Bigr].
\end{equation}
Unlike Eq.~\ref{h`-psc_ap}, the improved Hamiltonian in Eq.~\ref{h-psc_ap} is always 2-local regardless of the number of amino acids or the dimensionality of the
problem, since $d^{2}_{PQ}(N, D)$ is always 2-local.

First, notice that for valid configurations, all $(N-1)$ terms in the sum will
equal one, and $H_{psc}(N,D)$ evaluates to zero. If any of the
$d^{2}_{m,m+1}$ terms is zero, meaning that two amino acids occupy the same location, then $H_{onsite}$ will be drastically raised by the energy penalty $\lambda_0$. This can be achieved by setting $\lambda_0 >
\lambda_1$, and $\lambda_1 = N$. After excluding configurations in which any $d^{2}_{m,m+1}$
are zero, only configurations with values of $d^{2}_{m,m+1}>1$ are left. In
these instances, $H_{psc}(N,D)>0$ and $\lambda_1$ will play the role of an
energy penalty since $\lambda_1 >0$. Choosing $\lambda_1 = N$ and $\lambda_0 = N+1 > \lambda_1$ constrains unwanted or penalized configurations to eigenstates of $H_{protein}$ with
energies greater than zero, while plausible configurations of the protein correspond to energies less than or equal to zero. Note that the minimum energy of the HP model, in the case of all hydrophobic sequences with the maximum number of favorable contacts, is always greater than $-N$. This is satisfied in general for $N$ amino acids in either two or three dimensions.

\subsubsection{Pairwise hydrophobic interaction term, $H_{pairwise}$}\label{subsubsec:pairwise}
The HP model favors hydrophobic interactions by lowering the energy by
1 whenever non-nearest neighboring hydrophobic amino acids are a rectilinear
distance of 1 away.

This kind of interaction is represented by the following general expression:
\begin{equation} \label{h-pairw_ap}
H_{pairwise}(N,D) = -\sum_{i=1}^{N} \sum_{j=1}^{N} G_{ij}H^{ij}_{pairwise}
\end{equation}
Here $G$ is an $N \times N$ symmetric matrix with entries $G_{ij}$ equal to +1 when amino acids $i$ and $j$ are hydrophobic and non-nearest neighbors, and 0 otherwise. Note that $G_{ij}$ is set to zero for amino acids that are neighbors in the protein sequence. Notice also that alternate definitions of $G_{ij}$ could allow us to define lattice protein models that are more complex than the HP model. One of these models is the more realistic Miyazawa-Jernigan model \cite{Li2002} which includes interactions between 20 types of amino acids.

The form of $H^{ij}_{pairwise}$ depends on the spatial dimensionality of the problem. In two dimensions, we have
\begin{eqnarray}\label{h-pairw-ij-d2_ap}
H^{ij}_{pairwise} &=& H^{ij,2D}_{pairwise}(N) = x^{ij,2D}_{+}(N)+x^{ij,2D}_{-}(N) \nonumber\\&& +y^{ij,2D}_{+}(N)+y^{ij,2D}_{-}(N)
\end{eqnarray}
and in three dimensions,
\begin{eqnarray} \label{h-pairw-ij-d3_ap}
H^{ij}_{pairwise} = H^{ij,3D}_{pairwise}(N) = x^{ij,3D}_{+}(N)+x^{ij,3D}_{-}(N) \nonumber\\ +y^{ij,3D}_{+}(N) + y^{ij,3D}_{-}(N) + z^{ij,3D}_{+}(N)+z^{ij,3D}_{-}(N)
\end{eqnarray}

The terms on the right hand side of Eq.~\ref{h-pairw-ij-d3_ap} are independent; each one serves to query whether the $j$-th amino acid is located, with respect with the $i$-th amino acid, to the right, left, above, below, in front, or behind as represented by $x^{ij,3D}_{+}$, $x^{ij,3D}_{-}$, $y^{ij,3D}_{+}$, $y^{ij,3D}_{-}$, $z^{ij,3D}_{+}$, and $z^{ij,3D}_{-}$ terms, respectively. If the $j$-th amino acid is located at a distance of exactly one in any direction, $H^{ij}_{pairwise}$ is set to $+1$; otherwise it is set to zero. There
is a subtle but important condition embedded in these terms: they all vanish if the rightmost binary variable describing the $i$-th
residue's coordinate of interest (say $x$ for $x^{ij,3D}_{+}$ and
$x^{ij,3D}_{-}$ or $y$ for $y^{ij,3D}_{+}$ and $y^{ij,3D}_{-}$ or $z$ for
$z^{ij,3D}_{+}$ and $z^{ij,3D}_{-}$) does not end in 0, i.e., the coordinate has
to correspond to an even number. This is why we intentionally double count each pair of amino acids in Eq.~\ref{h-pairw_ap} by allowing both indexes $i$ and $j$ iterate from 1
to $N$. No special treatment is provided for the case where $i=j$,
since the diagonal terms of $G_{ij}$ are all zero due to the lack of amino acid
self interaction. Finally, because we want the interaction to be attractive
rather than repulsive, we use the minus sign in Eq.~\ref{h-pairw_ap}.

\textbf{The case of $N$ amino acids in a two dimensional grid for $N=2^M$ and
  $M\ge3$:} The terms listed below correspond to the pairwise interaction
Hamiltonian terms described above. The expressions below were constructed for $M
\ge 3$. The four amino acid case ($M=2$) is much simpler and will be discussed in
Sec.~\ref{sec:n4d2}. The expression for $x^{ij,2D}_{+}(N)$ is

\begin{eqnarray}\label{x_ij2D_+}
x^{ij,2D}_{+}(N) = (1 - q_{f(i,1)+1}) q_{f(j,1)+1} \prod_{s=2}^{\log_2
  N}(1-q_{f(j,1)+s} \nonumber\\ - q_{f(i,1)+s} + 2\,q_{f(j,1)+s}\, q_{f(i,1)+s})
\prod_{r=1}^{\log_2 N} (1  - q_{f(i,2)+r} \nonumber\\ - q_{f(j,2)+r}
+2q_{f(i,2)+r}q_{f(j,2)+r})\end{eqnarray}

The first two factors of $x^{ij,2D}_{+}(N)$ (Eq.~\ref{x_ij2D_+}) treat the
rightmost binary digit of the $x$ position of the $i$-th and $j$-th amino acid.
The first factor guarantees that the $i$-th residue is in an even position on the $x$-axis. For an interaction to be considered, the position of the $j$-th residue on the $x$-axis must be odd, as required by the second factor 
$q_{f(j,1)+1}$. The remaining factors of $x^{ij,2D}_{+}$ are {\sc xnor} functions that ensure that the rest of the binary digits that encode the $x$ position are equal for the $i$-th
and $j$-th amino acids. Finally, all the digits encoding the $y$ position have to
be equal, so that the $i$-th and $j$-th amino acids are nearest neighbors displaced only in the $x$-directionforcing the two residues to be in the same row. If all these
conditions are satisfied, $x^{ij,2D}_{+}$ evaluates to +1; otherwise it evaluates to 0.
These conditions rely on the fact that adding 1 to an even number only changes
the rightmost binary digit from 0 to 1.

The construction of  $y^{ij,2D}_{+}$ follows the same procedure as that of
$x^{ij,2D}_{+}$, namely,

\begin{eqnarray}\label{y_ij2D_+}
y^{ij,2D}_{+}(N) = (1 - q_{f(i,2)+1})q_{f(j,2)+1} \prod_{s=2}^{\log_2 N}(1-q_{f(j,2)+s} \nonumber\\ - q_{f(i,2)+s} + 2\,q_{f(j,2)+s}\, q_{f(i,2)+s})\prod_{r=1}^{\log_2 N} (1 - q_{f(i,1)+r} \nonumber\\- q_{f(j,1)+r} + 2q_{f(i,1)+r}q_{f(j,1)+r})
\end{eqnarray}

\begin{figure}[h]
\begin{center}
 \includegraphics[scale=0.5]{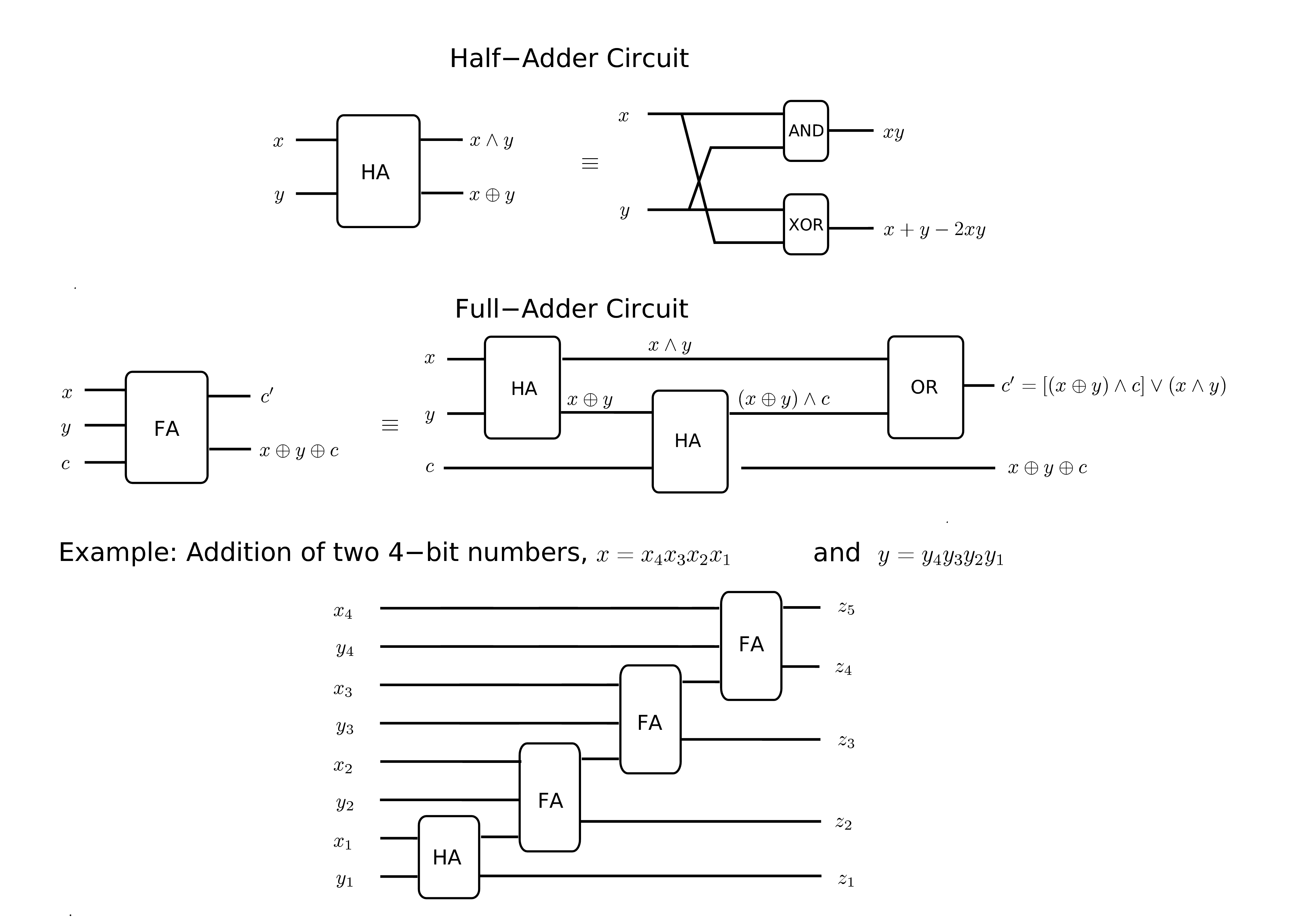}

 \caption{\label{fig:alg_circuit} Half-adder and full-adder components for the
   addition circuit implemented in the pairwise interaction Hamiltonian. We show
   the implementation of these two components for the addition of two
   4-bit numbers yielding $z=z_{5} z_{4} z_{3} z_{2} z_{1}$. The addition of
   $n$-bit numbers can be generalized trivially.}
 \end{center}
\end{figure}

The construction of $x^{ij,2D}_{-}$,
\begin{eqnarray} \label{x_ij2D_-}
x^{ij,2D}_{-}(N) =  (1 - q_{f(i,1)+1})q_{f(j,1)+1} \Bigl[1-\prod_{k=1}^{\log_2 N}(1 - \nonumber\\
q_{f(i,1)+k})\Bigr] (q_{f(j,1)+2}+q_{f(i,1)+2} -  2\,q_{f(j,1)+2}\, q_{f(i,1)+2}) \nonumber\\
\prod_{r=3}^{\log_2 N} \Bigl[1 - (q_{f(j,1)+r} + \prod_{u=2}^{r-1} q_{f(j,1)+u} - 2\prod_{u=2}^{r} q_{f(j,1)+u})\nonumber\\
 - q_{f(i,1)+r} + 2 q_{f(i,1)+r}(q_{f(j,1)+r} + \nonumber\\
\prod_{u=2}^{r-1} q_{f(j,1)+u} - 2 \prod_{u=2}^{r} q_{f(j,1)+u})\Bigr] \nonumber\\
\prod_{s=1}^{\log_2 N} (1 - q_{f(i,2)+s} - q_{f(j,2)+s} + 2q_{f(i,2)+s}q_{f(j,2)+s})
\end{eqnarray}
involves several considerations. As in the expression for $x^{ij,2D}_{+}$, the first factor $(1 - q_{f(i,1)+1})$ tests if the $i$-th amino acid is in an even position along the $x$-axis. Here, we are interested in querying whether the $j$-th amino acid is directly to the left of the $i$-th, and apply a different procedure than that of Eq.~\ref{x_ij2D_+}. We add $00 \cdots 01$ to the $x$ coordinate of the $j$-th residue, thus moving ``right'' by one unit, and use the {\sc xnor} function to check if the result matches the $x$ coordinate of the $i$-th amino acid. The problem is not as trivial as the case of $x^{ij,2D}_{+}$. Setting $i$ at an even coordinate value along the axis of interest forces $j$ to be in an odd coordinate. However, adding $00 \cdots 01$ to an odd binary number in general will change more digits than just the last digit due to carry bits. We used the circuit presented in Fig.~\ref{fig:alg_circuit} and the Boolean algebra introduced in Sec.~\ref{sec:comp_alg} to obtain the general expression for the addition of $00 \cdots 01$ to an $n$-bit number. If we take $x= x_{n} x_{n-1} \cdots x_{2} x_{1}$ and $y=00 \cdots 01$, then the result $z=z_{n+1} z_{n} z_{n-1} \cdots z_{2} z_{1}$ for the addition $z=x+y$ is
the recursive algebraic expression,
\begin{align*}
z_{1} &= 0\\
z_{2} &= 1-x_{2}\\
z_{k} &= x_{k} + \prod_{u=2}^{k-1} x_{u} - 2 \prod_{u=2}^{k} x_{u} \quad \textrm{for} \quad 3 \le k \le n \\
z_{n+1} &= \prod_{u=2}^{n} x_{u}
\end{align*}

As in the case of $x^{ij,2D}_{+}$, we impose
conditions that guarantee that the $y$ coordinate is the same for both amino
acids (that they are in the same row).

A special case arises when the $j$-th amino acid is at the rightmost position in the grid, with an $x$ coordinate value of $11 \cdots 11$. When $00 \cdots 01$ is added to this coordinate, $z_{n+1}$ evaluates to 1 and the $n$ bits $z_1$ to $z_n$ evaluate to 0. Since only the first $n$ bits are used to compare coordinates, this $z$ would be an undesirable match with an $i$-th amino acid positioned at $x = 00 \cdots 00$. Notice that a value of $x = 00 \cdots 00$ positions the $i$-th amino acid positioned at the minimal/leftmost position in the grid, for which $x^{ij,2D}_{-}$ should not
even be considered. The factor $[1-\prod_{k=1}^{\log_2 N}(1 - q_{f(i,1)+k})]$ in
Eq.~\ref{x_ij2D_-} sets the term $x^{ij,2D}_{-}$ to 0 if the $x$ coordinate of
the $i$-th amino acid is $00 \cdots 00$, taking care of both of these concerns. 

The construction of  $y^{ij,2D}_{-}$ follows the same procedure as that of
$x^{ij,2D}_{-}$, namely,
\begin{eqnarray} \label{y_ij2D_-}
y^{ij,2D}_{-}(N) =  (1 - q_{f(i,2)+1})q_{f(j,2)+1} \Bigl[1-\prod_{k=1}^{\log_2 N}(1 - \nonumber\\
q_{f(i,2)+k})\Bigr] (q_{f(j,2)+2}+q_{f(i,2)+2} - 2\,q_{f(j,2)+2}\, q_{f(i,2)+2}) \nonumber\\
\prod_{r=3}^{\log_2 N} \Bigl[1 - (q_{f(j,2)+r} + \prod_{u=2}^{r-1} q_{f(j,2)+u} - 2 \prod_{u=2}^{r} q_{f(j,2)+u})\nonumber\\
 - q_{f(i,2)+r} + 2 q_{f(i,2)+r} (q_{f(j,2)+r} + \nonumber\\
\prod_{u=2}^{r-1} q_{f(j,2)+u} - 2 \prod_{u=2}^{r} q_{f(j,2)+u})\Bigr] \nonumber\\
\prod_{s=1}^{\log_2 N} (1 -q_{f(i,1)+s} - q_{f(j,1)+s} + 2q_{f(i,1)+s}q_{f(j,1)+s})
\end{eqnarray}

The three-dimensional extension of these equations is presented in the Appendix.


\subsection{Maximum locality and scaling of the number of terms in $H_{protein}$}

In this section, we estimate the number of terms included in the total
Hamiltonian $H_{protein}$ and present procedures required to reduce the locality
of the terms to 2-local. These estimates assess the size of a
quantum device necessary for eventual experimental realizations of the
algorithm. The reduction of the locality of the terms involves ancillary qubits.

Each amino acid requires $D \log_2 N$ qubits to specify its position in the
lattice. Since our algorithm fixes the position of two amino acids, the
number of qubits needed to encode the coordinates of the $(N-2)$ remaining amino
acids is $(N-2)D \log_2 N$. From the expressions given for $H_{onsite}$,
$H_{psc}$ and $H_{pairwise}$, one can deduce that the maximum locality is
determined by $2D\log_2 N$ --- the number of qubits corresponding to two amino
acids. As described in Sec.~\ref{sec:psc}, the $H_{psc}$ term is always 2-local
in nature regardless of the number of amino acids. For scaling arguments, it is
crucial to point out that all possible 1-local and 2-local terms, that account for ${(N-2)D \log_2 N}$
and $\binom{(N-2)D \log_2 N}{2}$ total terms, repectively, appear in the
expansion, but that not all possible 3-local or higher locality terms will be
present. For example, the terms $q_i q_j q_k$, where the indexes $i$, $j$ and
$k$ are associated with three different amino acids, are not part of the
expansion, since every term should only involve products of qubits describing two amino acids, regardless of its locality. Table~\ref{table:locality}
summarizes the number of $k$-local terms required to construct the protein
Hamiltonian, $H_{protein}=H_{onsite} + H_{psc} + H_{pairwise}$. The alternative
count from the combinatorial expressions of Table~\ref{table:locality} scales
as $N^6$ for $D=2$ and as $N^8$ for $D=3$. Table~\ref{table:locality} provides the exact term count.

\begin{table}[h]
 \caption{\label{table:locality} The number of $k$-local terms obtained in the final expression for $H_{protein}$ as a function of the number of amino acids $N$, $N = 2^M$, and dimensions ($D$) of the lattice.}
\begin{ruledtabular}
\begin{tabular}{c|c}
locality  & Number of terms, $T_k$ \\ \hline $k=0$  & 1 \\ $k=1$ & $(N-2) D \log_2 N$ \\ $2 \le k \le D \log_2 N$ & $\binom{N-2}{2} \sum_{i=1}^{k-1} \binom{D \log_2 N}{i} \binom{D \log_2 N}{k-i} + (N-2) \binom{D \log_2 N}{k}$\\ $D \log_2 N < k \le 2 D \log_2 N$  & $\binom{N-2}{2} \sum_{i=k-D \log_2 N}^{D \log_2 N} \binom{D \log_2 N}{i} \binom{D \log_2 N}{k-i}$  \\ Total number of terms & $\sum_{k=0}^{2D log_2 N} T_k \sim N^{2D+2}$
\end{tabular}
\end{ruledtabular}
\end{table}

\section{Case study: HPPH}\label{sec:n4d2}
With the goal of designing an experiment for adiabatic quantum computers with small numbers of qubits, we
concentrate on the simplest possible instance of the HP-model -- a four amino acid
loop that contains a favorable interaction and therefore ``folds''. 

In Sec.~\ref{sec:h-n4d2} we present the protein Hamiltonian, followed by the
partitioning of the $N$-local Hamiltonian terms to 2-local. Finally, we present
numerical simulations which confirm the local minimum through the use of the proposed algorithm.

\subsection{Hamiltonian terms for the case of four amino acids in 2D}\label{sec:h-n4d2}

The onsite Hamiltonian for this example takes the form
\subsubsection{Onsite term, $H_{onsite}$}
\begin{eqnarray}\label{h-onsite-n4d2}
H_{onsite}(N=4,D=2) = \lambda_0 (H_{onsite}^{12}+H_{onsite}^{13}+ \nonumber\\
H_{onsite}^{14}+H_{onsite}^{24}+H_{onsite}^{34})
 \end{eqnarray}
with
\begin{eqnarray}\label{h-onsite-ij-n4d2}
H_{onsite}^{ij}(N=4,D=2) = \prod_{k=1}^{2} \prod_{r=1}^{2} \Bigl(1-q_{f(i,k)+r}- \nonumber\\
q_{f(j,k)+r}+2\,q_{f(i,k)+r}\,q_{f(j,k)+r}\Bigr)
 \end{eqnarray}
and
\begin{equation}
f(i,k)=4(i-1)+2(k-1).
\end{equation}
Note that $H_{onsite}^{23}$ does not appear in Eq.~\ref{h-onsite-n4d2} since, as
described in Sec.~\ref{subsec:mapping}, the two central amino
acids are fixed in position and guaranteed not to occupy overlapping gridpoints that would contribute an energy penalty to the onsite
term  {\it a priori} . On the other hand, other terms involving amino acids 2 and 3 cannot be discarded, since these amino acids will affect their other neighbors through $H_{psc}$ and they can participate in hydrophobic interactions through $H_{pairwise}$.

\subsubsection{Primary structure constraint term, $H_{psc}$}\label{sec:psc-n4d2}
The pairwise term
\begin{equation}
d^{2}_{PQ}(N=4, D=2) = \sum_{k=1}^{2} \Bigl(\sum_{r=1}^{2} 2^{r-1}(q_{f(P,k)+r}-q_{f(Q,k)+r}) \Bigr)^2
\end{equation}
with
\begin{eqnarray}
H_{psc}(N=4,D=2) &=& \lambda_1 \bigl(-3 + d^{2}_{12}+ d^{2}_{23}+ d^{2}_{34}\bigr) \nonumber\\
&=& \lambda_1 \bigl(-2 + d^{2}_{12}+ d^{2}_{34}\bigr)
\end{eqnarray}
takes advantage of the fact that $d^{2}_{23}=1$ by construction.

\subsubsection{Pairwise term, $H_{pairwise}$}
Finally, a pairwise interaction term is required to impose an energy
stabilization for non-nearest neighbor hydrophobic amino acids that occupy adjacent sites in the lattice.

For the sequence HPPH,
\begin{equation}
 G = \begin{pmatrix}
      0 & 0 & 0 & 1 \\
      0 & 0 & 0 & 0 \\
      0 & 0 & 0 & 0 \\
      1 & 0 & 0 & 0 \\ 
     \end{pmatrix}
\end{equation}
and therefore,
\begin{equation} \label{h-pairw-d2}
H^{2D}_{pairwise}(N=4,D=2) = -(H^{14,2D}_{pairwise}+H^{41,2D}_{pairwise}).
\end{equation}
For this particular case of interest
\begin{eqnarray}\label{h-pairw-ij-d2}
H^{ij,2D}_{pairwise}(N=4) = x^{ij,2D}_{+}(N=4)+x^{ij,2D}_{-}(N=4)+ \nonumber\\
y^{ij,2D}_{+}(N=4)+y^{ij,2D}_{-}(N=4).
\end{eqnarray}

\begin{figure}[h]
 \begin{center}
  \includegraphics[scale=0.60]{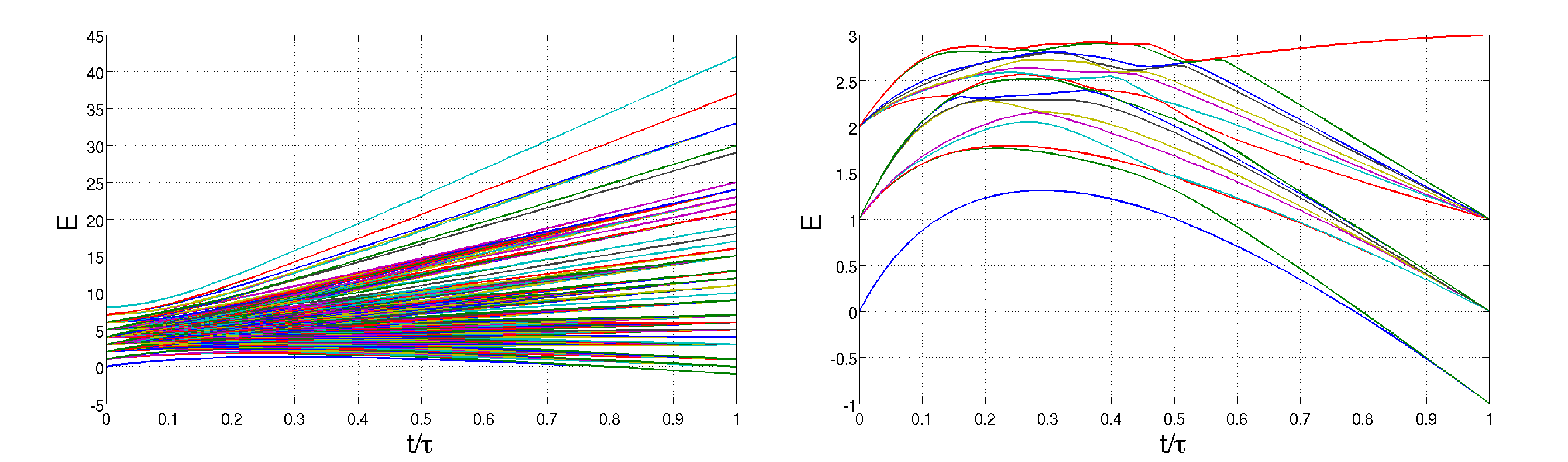}
  \caption{\label{fig:hpph} (Color online) Spectrum of the instantaneous energy eigenvalues for
    the 8-local time dependent Hamiltonian used in the algorithm for the peptide
    HPPH (left). The plot to the right examines the lowest 15 states of the 256
    states from the left.}
  \end{center}
\end{figure}

The explicit forms of these functions are:
\begin{eqnarray}\label{xplus-n4d2}
x^{ij,2D}_{+}(N=4) = (1 - q_{f(i,1)+1}) q_{f(j,1)+1} (1-q_{f(j,1)+2}-\nonumber\\
q_{f(i,1)+2} + 2\,q_{f(j,1)+2} q_{f(i,1)+2}) \nonumber\\
\prod_{s=1}^{2} (1 - q_{f(i,2)+s} - q_{f(j,2)+s} + 2q_{f(i,2)+s}q_{f(j,2)+s}),
\end{eqnarray}
\begin{eqnarray}\label{yplus-n4d2}
y^{ij,2D}_{+}(N=4) = (1 - q_{f(i,2)+1}) q_{f(j,2)+1} (1-q_{f(j,2)+2}- \nonumber\\
q_{f(i,2)+2} + 2\,q_{f(j,2)+2}\, q_{f(i,2)+s}) \nonumber\\
\prod_{s=1}^{2} (1 - q_{f(i,1)+s} - q_{f(j,1)+s} + 2q_{f(i,1)+s}q_{f(j,1)+s}),
\end{eqnarray}
\begin{eqnarray}\label{xminus-n4d2}
x^{ij,2D}_{-}(N=4) = (1 - q_{f(i,1)+1}) q_{f(j,1)+1} q_{f(i,1)+2} \nonumber\\
(q_{f(j,1)+2}+q_{f(i,1)+2} - 2 q_{f(j,1)+2} q_{f(i,1)+2})\prod_{s=1}^{2} (1 - \nonumber\\
q_{f(i,2)+s} - q_{f(j,2)+s} + 2\,q_{f(i,2)+s}q_{f(j,2)+s}),
\end{eqnarray}
\begin{eqnarray}\label{yminus-n4d2}
y^{ij,2D}_{-}(N=4) = (1 - q_{f(i,2)+1})q_{f(j,2)+1}q_{f(i,2)+2} \nonumber\\
(q_{f(j,2)+2}+q_{f(i,2)+2} - 2\,q_{f(j,2)+2} q_{f(i,2)+2})\prod_{s=1}^{2} (1 - \nonumber\\
q_{f(i,1)+s} - q_{f(j,1)+s} + 2\,q_{f(i,1)+s}q_{f(j,1)+s}).
\end{eqnarray}
After expanding all of the terms in $H_{onsite}$, $H_{psc}$ and $H_{pairwise}$, we fix amino acids 2 and 3 as described in Sec.~\ref{subsec:mapping}, substituting the variables $q_{12} q_{11} q_{10} q_9\,\,q_8 q_7 q_6 q_5$ by the constant values $0110\,\,0101$ as shown in Fig.~\ref{fig:2Dgrid}. The final expression for $H_{protein}$ now depends on the 8 binary variables encoding
the coordinates of amino acids 1 and 4, $q_4 q_3 q_2 q_1$ and $q_{16} q_{15} q_{14} q_{13}$, respectively. For convenience in notation, we relabel the coordinates of amino acid 4 from $q_{16} q_{15} q_{14} q_{13}$ to $q_8 q_7 q_6 q_5$. After these substitutions, the final expression for the energy function $H_{protein}$ will be dependent on products involving the variables $q_1$ through $q_8$. Following the mapping explained at the end of Sec.~\ref{subsec:mapping}, the quantum expression for $\hat H_{protein}$ is a $2^8 \times 2^8$ matrix. This Hamiltonian matrix defines the final Hamiltonian $\hat H(t = \tau)$ of the adiabatic evolution. The initial Hamiltonian representing the transverse field whose ground
state is a linear superposition of all $2^8$ states in the computational basis
can be written as
\begin{equation}
 \hat H_0 \equiv \hat H(t=0) = \sum_{i=1}^{8} \hat q_x^{i} = \sum_{i=1}^{8} \frac{1}{2}(I-\hat{\sigma}^{x}_{i})
\end{equation}
with
\begin{equation}
 \ket{\psi_{g} (t=0)} = \frac{1}{\sqrt{2^{8}}} \sum_{q_{i} \in \{0,1\}} \ket{q_{8} q_{7} q_{6} q_{5} q_{4} q_{3} q_{2} q_{1}}
\end{equation}
Finally, we can construct a time dependent Hamiltonian as shown in Eq.~\ref{h(t)},
\begin{equation}
 \hat H(t) = (1-t/\tau) \hat H_0 + (t/\tau) \hat H_{protein}
\end{equation}
This time dependent Hamiltonian is also a $2^8 \times 2^8$ matrix as well. The instantaneous spectrum can be obtained by diagonalizing at every $t/\tau$ without need to specify $\tau$. Since $\tau$ is the running time, we are interested in $0 \le t/\tau \le 1$. The spectrum of the corresponding $\hat H(t)$ for this four amino acid peptide HPPH is given in Fig.~\ref{fig:hpph}.

\begin{turnpage}
\begin{figure*}[p]
\begin{center}
 \includegraphics[scale=0.70]{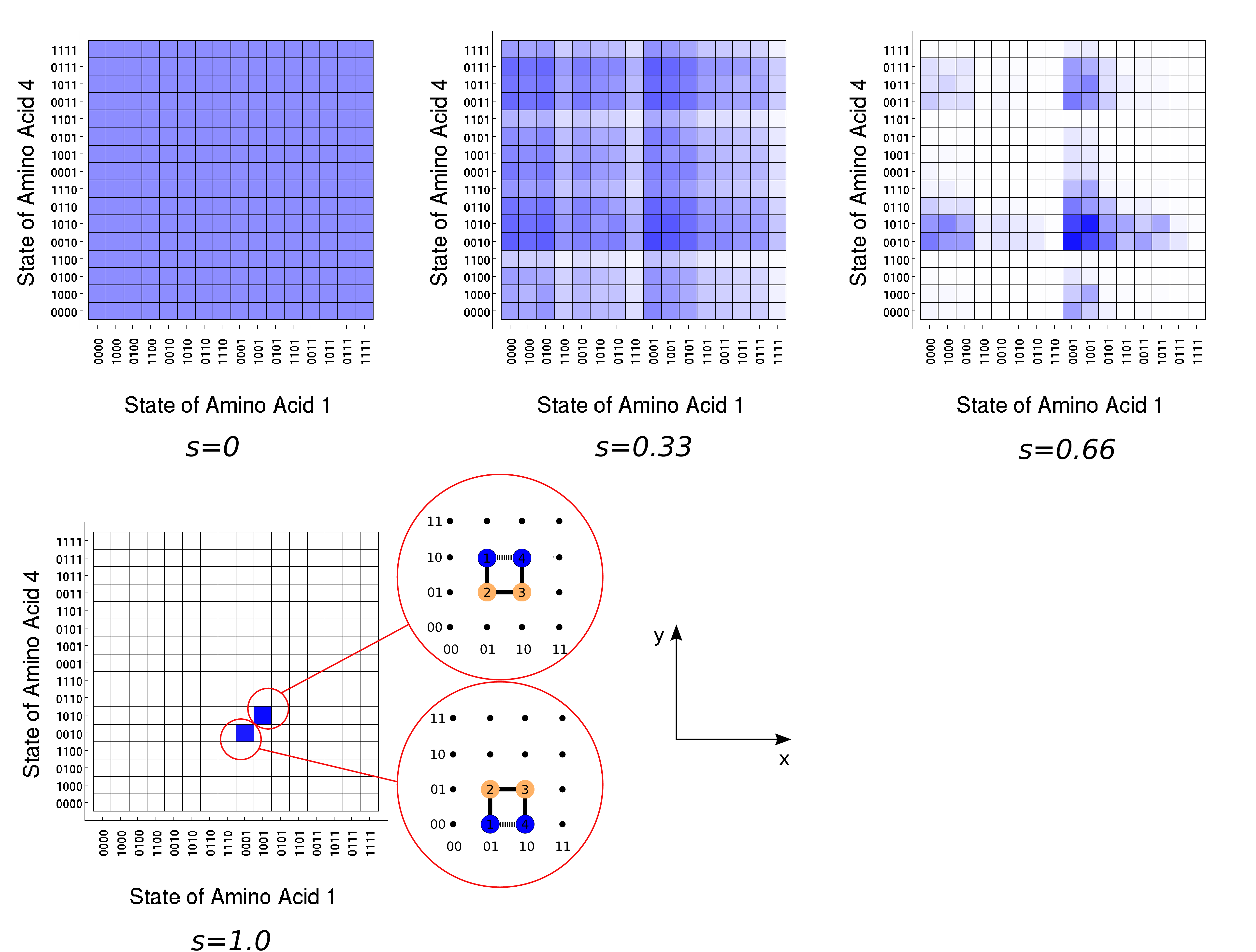}
 \caption{\label{fig:animation} (Color online) Snapshots of the instantaneous ground state for $H(t)$. The brightness of the box is proportional to $\vert c_n\vert^2$. Axis labels and state vectors for each particular box correspond to $\ket{\psi}=\sum_{n=0}^{255} c_n \ket{n}$ with $\ket{n}$ the $n$-th state vector out of the 256 possibilities given by $\ket{q_{16}} \ket{q_{15}} \ket{q_{14}} \ket{q_{13}} \ket{q_{4}} \ket{q_{3}} \ket{q_{2}} \ket{q_{1}}$. Notice that the $x$ axis is given by $\ket{q_{4}} \ket{q_{3}} \ket{q_{2}} \ket{q_{1}}$ and the $y$ axis given by $\ket{q_{16}} \ket{q_{15}} \ket{q_{14}} \ket{q_{13}}$. The final state corresponds to the two degenerate minima shown at the end}
 \end{center}
\end{figure*}
\end{turnpage}

Snapshots of the instantaneous ground state are shown in
Fig.~\ref{fig:animation}. Even though these snapshots do not correspond to
explicit propagation of the Schr\"{o}dinger equation, they indicate that
the final $H_{protein}$ is correct and that it provides the correct answer if a
sufficiently long time $\tau$ is allowed. Notice that at $t/\tau=0$, the
amplitude for all 256 states is equal, indicating a uniform superposition of all
states; at $t/\tau=1$, the readout corresponds to the two degenerate solutions
of HPPH.

\b{\section{Converting an N-local Hamiltonian to a 2-local Hamiltonian}\label{sec:2local}}

Motivated by the possibility of an experimental implementation, we explain how
to reduce the locality of a Hamiltonian from $k$-local to 2-local while conserving
its low-lying spectrum. We use Boolean reduction techniques \cite{boros02pseudoboolean,BIAMONTE2008} for Hamiltonians contructed from energy functions with structure similar to $H_{protein}$, where all of terms are sums of tensor products of $\sigma_{z}^{i}$ operators. By reducing the locality of the interactions, we introduce new ancilla qubits to represent higher order
interactions with sums of at most 2-local terms. Here, we present an illutrative example with a relative simple energy function but the methodology can be immediately extended to higher locality energy functions such as the one resulting in $H_{protein}$.

Consider a 4-local energy function of the form
\begin{equation}\label{h4-h2}
H_{toy}(q) = 1 +q_1 -q_2 + q_3 + q_4 - q_1 q_2 q_3 + q_1 q_2 q_3 q_4.
\end{equation}
As shown in Table~\ref{table:4localh}, this energy function has a unique minimum energy given by $q=q_4 q_3 q_2 q_1 = 0010$. The energy associated with this configuration is 0 in arbitrary units and all other possible values of the binary variables $q_1, q_2$, $q_3$ and $q_4$ have energies ranging from 0 to 4.

\begin{table}[h]
 \caption{\label{table:4localh} Truth table for the energy function $H_{toy}(q) = 1 +q_1 -q_2 + q_3 + q_4 - q_1 q_2 q_3 + q_1 q_2 q_3 q_4$.}
\begin{ruledtabular}
\begin{tabular}{cccc|c}
$q_4$ & $q_3$ & $q_2$ & $q_1$ & $H(q_1,q_2,q_3,q_4)$ \\ \hline 0 & 0 & 1 & 0 & 0 \\ 0 & 0 & 0 & 0 & 1 \\ 0 & 0 & 1 & 1 & 1 \\ 0 & 1 & 1 & 0 & 1 \\ 0 & 1 & 1 & 1 & 1 \\ 1 & 0 & 1 & 0 & 1 \\ 0 & 0 & 0 & 1 & 2 \\ 0 & 1 & 0 & 0 & 2 \\ 1 & 0 & 0 & 0 & 2 \\ 1 & 0 & 1 & 1 & 2 \\ 1 & 1 & 1 & 0 & 2 \\0 & 1 & 0 & 1 & 3 \\ 1 & 0 & 0 & 1 & 3 \\   1 & 1 & 0 & 0 & 3 \\ 1 & 1 & 1 & 1 & 3 \\ 1 & 1 & 0 & 1 & 4 \\
\end{tabular}
\end{ruledtabular}
\end{table}

The goal is to obtain an energy function $H'$ that preserves these energies along with their associated
bit strings, but defines $H'$ using only 1-local and 2-local terms. That is, the goal is to obtain a substitution for $H_{toy}$ with the following form,
\begin{equation}\label{decomp_map}
 H'(\tilde q_1, \cdots ,\tilde q_M) = c_0 + \sum_{i=1}^{M} c_i \tilde{q}_i + \sum_{i=1}^{M-1} \sum_{j=i+1}^{M} d_{ij}\tilde{q}_i \tilde{q}_j.
\end{equation}
In Eq.~\ref{decomp_map} the new set of binary variables $\tilde{q}$ includes the
original variables $q_i$ as well as ancillary variables required to reduce locality. The extra ancillary bits raise the total number of variables to $M$. 

Since the information contained within the problem and the solution we are seeking both rely on the original set of $q$ variables (in the case of protein folding, for example, the string $q$ encodes the positions of the amino acids in the lattice), we must be able to identify values corresponding to the original $q$, regardless of the substitutions made to convert a $k$-local function to a 2-local. The new energy function $H'$ needs to have the energy values of the original function in its energy spectrum. In addition, the values of the bit string $\tilde q$ for these energies must match the same values of $q$ in the original function. For the particular example of Eq.~\ref{h4-h2}, consider the substitutions, $q_1 q_2 \rightarrow \tilde q_5$ and $q_3 q_4 \rightarrow \tilde q_6$. These two subtitutions introduce two new independent binary variables, $\tilde q_5$ and $\tilde q_6$ and regardless of the values of $q_1, q_2, q_3$ and $q_4$, they can take any value in $\{0,1\}$. Since we want to preserve both the physical meaning of the original energy function, as well as its energy spectrum, we need to perform an action on the cases where the conditions $\tilde q_5 = q_1 \wedge q_2$ and $\tilde q_6 = q_3 \wedge q_4$ are not satisfied and lack any meaning in the context of the original energy function. One way to address this problem while keeping the original spectrum intact is to add a penalty function which enforces the conditions $\tilde q_5 = q_1 \wedge q_2$ and $\tilde q_6 = q_3 \wedge q_4$. For every substitution of the form $q_i q_j \rightarrow \tilde q_n$, consider a function of the form \cite{BIAMONTE2008}
\begin{equation}
 H_\wedge (q_i, q_j, \tilde q_n) = \delta (3 \tilde q_n + q_i q_j - 2 q_i \tilde q_n - 2 q_j \tilde q_n).
\end{equation}
As shown in Table~\ref{table:hwedge}, for $\delta > 0$, the function $H_\wedge (q_i, q_j, \tilde q_n)$ is greater than zero whenever $\tilde q_n \neq q_i \wedge q_j$ and it evaluates to zero whenever $\tilde q_n = q_i \wedge q_j$.

\begin{table}[h]
 \caption{\label{table:hwedge} Truth table for the function $H_{\wedge}(q_i,q_j,\tilde q_n) = \delta (3 \tilde q_n + q_i q_j - 2 q_i \tilde q_n - 2 q_j \tilde q_n)$ used for the locality reduction procedure described in Sec.~\ref{sec:2local}.}
\begin{ruledtabular}
\begin{tabular}{ccc|c}
$\tilde q_n$ & $q_i$ & $q_j$ & $H_{\wedge}(q_i,q_j,\tilde q_n)$ \\ \hline 0 & 0 & 0 & 0\\ 0 & 0 & 1 & 0 \\ 0 & 1 & 0 & 0 \\ 1 & 1 & 1 & 0 \\ \hline 1 & 0 & 0 & $3 \delta$\\ 1 & 0 & 1 & $\delta$ \\ 1 & 1 & 0 & $\delta$ \\ 0 & 1 & 1 & $\delta$ \\
\end{tabular}
\end{ruledtabular}
\end{table}

A two-local expression of the form presented in Eq.~\ref{decomp_map} can be obtained by adding one $H_\wedge (q_i, q_j, \tilde q_n)$ function for each substitution $q_1 q_2 \rightarrow \tilde q_5$ and $q_3 q_4 \rightarrow \tilde q_6$ and by making the additional trivial substitutions $q_1 \rightarrow \tilde q_1$, $q_2 \rightarrow \tilde q_2$, $q_3 \rightarrow \tilde q_3$, and $q_4 \rightarrow \tilde q_4$, to conveniently change in notation to the set of binary variables $\tilde q$ . For the case of the energy function of Eq.~\ref{h4-h2}, the locality reduced version is
\begin{align}\label{eq:htoyred}
 H_{toy,reduced}(\tilde q) &= 1 + \tilde q_1 - \tilde q_2 + \tilde q_3 + \tilde q_4 -\tilde q_5 \tilde q_3 + \tilde q_5 \tilde q_6 + H_\wedge (q_1, q_2, \tilde q_5) + H_\wedge (q_3, q_4, \tilde q_6)\nonumber\\
& =  1 + \tilde q_1 - \tilde q_2 + \tilde q_3 + \tilde q_4 -\tilde q_5 \tilde q_3 + \tilde q_5 \tilde q_6 +  \delta (3 \tilde q_5 + \tilde q_1 \tilde q_2 - 2 \tilde q_1 \tilde q_5 - 2 \tilde q_2 \tilde q_5) \nonumber\\
& +\delta (3 \tilde q_6 + \tilde q_3 \tilde q_4 - 2 \tilde q_3 \tilde q_6 - 2 \tilde q_4 \tilde q_6).
\end{align}
Recall that the additional functions $H_\wedge (\tilde q_1, \tilde q_2, \tilde q_5)$ and $H_\wedge (\tilde q_3, \tilde q_4, \tilde q_6)$ increase the energy of $H_{toy,reduced}$ by at least $\delta$ whenever the conditions $\tilde q_5 = \tilde q_1 \wedge \tilde q_2$ and $\tilde q_6 = \tilde q_3 \wedge q_4$ are not satisfied. Table~\ref{table:2localh} shows the one-to-one mapping between the energies of non-penalized configurations of $H_{toy,reduced}(\tilde q)$ and configurations presented in Table~\ref{table:4localh} associated with $H_{toy}(q)$. Even though there is a unique configuration $\{\tilde q_6 = \tilde q_3 \wedge \tilde q_4, \tilde q_5 = \tilde q_1 \wedge \tilde q_2, \tilde q_4, \tilde q_3, \tilde q_2, \tilde q_1\}$ associated with every $\{q_1, q_2, q_3, q_4\}$ with the same energy, it does not necessarily hold that the lowest $2^4$ out of the $2^6$ energies of $H_{toy,reduced}$ consist of the $2^4$ energies of $H_{toy}$. For example, if we pick a small penalty $\delta$ in Table~\ref{table:2localh}, say $0 \le \delta \le 4$, then some of the states penalized by either $H_\wedge (\tilde q_1, \tilde q_2, \tilde q_5)$ or $H_\wedge (\tilde q_3, \tilde q_4, \tilde q_6)$ can still have an energy within the energy values of $H_{toy}$. To avoid this situation, we can choose $\delta > \max (H_{toy})$ which will be sufficient to remove the energies of the penalized states from the region corresponding to energies of $H_{toy}$, therefore conserving the low-lying spectra of the original $H_{toy}$. Using the mapping explained at the end of Sec.~\ref{subsec:mapping}, the quantum version of the 4-local energy function from Eq.~\ref{h4-h2} is:
\begin{equation}\label{eq:htoyquantum}
\hat H_{toy} = I +\hat{\tilde q}_1 -\hat{\tilde q}_2 + \hat{\tilde q}_3 + \hat{\tilde q}_4 - \hat{\tilde q}_1 \hat{\tilde q}_2 \hat{\tilde q}_3 + \hat{\tilde q}_1 \hat{\tilde q}_2 \hat{\tilde q}_3 \hat{\tilde q}_4.
\end{equation}
The quantum version of the 2-local reduced form presented in Eq.~\ref{eq:htoyred} is,
\begin{align}\label{eq:quantumhtoyred}
 \hat H_{toy,reduced} &= I + \hat{\tilde q}_1 - \hat{\tilde q}_2 + \hat{\tilde q}_3 + \hat{\tilde q}_4 -\hat{\tilde q}_5 \hat{\tilde q}_3 + \hat{\tilde q}_5 \hat{\tilde q}_6 + \delta (3 \hat{\tilde q}_5 + \hat{\tilde q}_1 \hat{\tilde q}_2 - 2 \hat{\tilde q}_1 \hat{\tilde q}_5 - 2 \hat{\tilde q}_2 \hat{\tilde q}_5) \nonumber\\
& +\delta (3 \hat{\tilde q}_6 + \hat{\tilde q}_3 \hat{\tilde q}_4 - 2 \hat{\tilde q}_3 \hat{\tilde q}_6 - 2 \hat{\tilde q}_4 \hat{\tilde q}_6) 
\end{align}
Notice that $\hat H_{toy}$ acts on a $2^4$ dimensional Hilbert space, span$\{\ket{\tilde q_4} \otimes \ket{\tilde q_3} \otimes \ket{\tilde q_2} \otimes \ket{\tilde q_1}\}$, while $\hat H_{toy,reduced}$ acts on a $2^6$ dimensional Hilbert space, span$\{\ket{\tilde q_6} \otimes \ket{\tilde q_5} \otimes \ket{\tilde q_4} \otimes \ket{\tilde q_3} \otimes \ket{\tilde q_2} \otimes \ket{\tilde q_1}\}$.

Due to the conservation of the spectrum and bit strings described above (as reflected in Tables~\ref{table:4localh} and \ref{table:2localh}), the solution obtained from an adiabatic quantum algorithm using either $\hat H_{toy}$ or $\hat H_{toy,reduced}$ as $\hat H_{final}$, 
\begin{equation}\label{htoy(t)}
 \hat H(t) = (1-t/\tau) \hat H(0) + (t/\tau) \hat H_{final}
\end{equation}
should be the same.

In the case of the 2-local Hamiltonian $\hat H_{toy,reduced}$, the solution to the optimization problem is obtained using an adiabatic algorithm after reading the qubits associated to $\tilde q_{4}, \tilde q_{3}, \tilde q_{2}, \tilde q_{1}$ at $t = \tau$ from the space span$\{\ket{\tilde q_6} \otimes \ket{\tilde q_5} \otimes \ket{\tilde q_4} \otimes \ket{\tilde q_3} \otimes \ket{\tilde q_2} \otimes \ket{\tilde q_1}\}$ at $t = \tau$. Notice that the ancillary qubits in the six qubit version do not carry any physical information, as expected, since all of the valuable information was stored in the qubits coming from the original expression before the reduction. The cost of reducing the locality of a Hamiltonian to another which contains at most two-body interactions is the increase in the number of resources due to the additional ancillary bits.

\begin{figure}[h]
\begin{center}
 \includegraphics[scale=1.00]{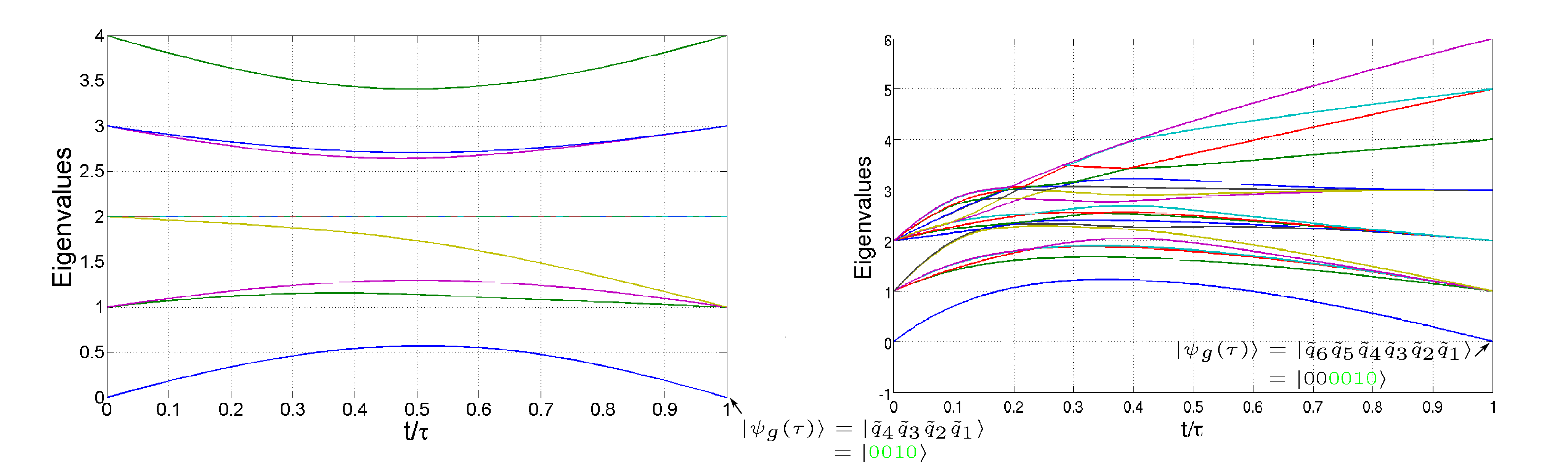}
 \caption{\label{fig:toy} (Color online) Spectrum comparison of the instantaneous energy eigenvalues for the 4-local toy Hamiltonian $\hat H_{toy}$ (left) and its corresponding 2-local version $\hat H_{toy,reduced}$(right). (left) Full spectrum of the $2^4$ instantaneous eigenvalues for $\hat H_{toy}(\tilde q_, \tilde q_2, \tilde q_3, \tilde q_4)$.  (right) First 19 instantaneous eigenvalues for the 2-local version of $\hat H_{toy}$, denoted as $\hat H_{toy,reduced}$ in text. The value used for $\delta$ is 5. The first $2^4$ levels, $0 \le$ eigenvalues $\le 4$, are associated to the original levels from $\hat H_{toy}$. The three remaining states with eigenvalues greater than 4 are penalized states which violate the conditions $\tilde q_n = \tilde q_i \wedge \tilde q_j$ (see Table~\ref{table:2localh} for details)}
 \end{center}
\end{figure}

 Figure~\ref{fig:toy} shows the the eigenenergies of Eq.~\ref{htoy(t)} vs. $t/\tau$, where $\hat H_{final}$ is replaced by $\hat H_{toy}$ (see Figure~\ref{fig:toy}(a)), and by $\hat H_{toy,reduced}$ with $\delta = 5$, (see Fig.~\ref{fig:toy}(b)). As expected from Table~\ref{table:4localh} and \ref{table:2localh}, Fig.~\ref{fig:toy} illustrates the preservation of the subsystem corresponding to the variables $\tilde q_1, \tilde q_2, \tilde q_3$ and $\tilde q_4$ in the ground state of both the original and reduced-locality Hamiltonian. Degeneracy and overlap of lines in the spectra in Fig.~\ref{fig:toy} make it difficult to graphically convey that both spectra in Fig.~\ref{fig:toy} indeed have 16 states for $0 \le$ eigenenergies $\le 4$. In Fig.~\ref{fig:toy}(b) we plotted the first 19 eigenstates out of the $2^6$ eigenstates corresponding to $\hat H_{toy,reduced}$. At $t/\tau = 1$, states with energy greater than 4 correspond to states which violate the {\sc and} condition introduced by the reduction process. Notice that there are two eigenstates with eigenvalue 5 in agreement with the table presented in Appendix~\ref{appendix:reduction} after substituting $\delta = 5$, and one state which corresponds to the one of the four-degenerate manifold with $E = 6$.

\section{Resources needed for a 2-local Hamiltonian expression in protein folding}\label{sec:resources}

For any $k$-local energy function, e.g., $h = q_1 q_2 \cdots q_k$, the
reduction can be carried out iteratively, adding the penalty function $H_\wedge
(q_i, q_j, \tilde q_n)$ for every substitution of the form $q_i q_j \rightarrow \tilde q_n$. For a
$k$-local term, $(k-2)$ substitutions are required for the reduction to 2-local,
and therefore require $(k-2)$ ancillary bits.

In the particular case of the protein Hamiltonian the reduction procedure needs
to be repeated $(N-2)(N^{D}-D \log_2 N -1)$ times, as described below. All the terms in the HP
Hamiltonian include among at most interactions two amino acids, which results
in a maximum locality of $2D\log_2 N$. In the following discussion, the cluster
notation $[k][l]$ specifies the contributions of a particular $(k+l)-$local term
into $k$ variable coming from an amino acid with index $i$ and $l$ variables from an amino acid with index $j$. Since all the terms are of this form, to obtain a 2-local Hamiltonian, all products corresponding to each $[k]$ and $[l]$ of each cluster have to be converted to 1-local terms. We reduce terms for variables describing each amino
acid in turn, for a total of $D\log_2 N$ variables. All possible combinations of two
variables from the $D\log_2 N$ variables for an amino acid are substituted. The
number of ancillary bits required for this substitution is $\binom{D \log_2
  N}{2}$. These substitutions convert all terms of the form $[3][0]$ and
$[2][1]$ to 2-local. To convert terms of the form $[4][0]$ or $[3][1]$ to 2-local
we need to consider $\binom{D \log_2 N}{3}$ terms originally containing three
variables from one amino acid. After employing an additional ancillary bit per term
and applying the previous reduction step, all these terms collapse to 1-local
with respect to the $i$-th amino acid, i.e., these terms will assume the form
$[1][l]$. Iterating over the $D \log_2 N$ variables for a specific amino acid in order of increasing locality will give us the
number of substitutions or ancilla bits needed per amino acid in order to reduce a particular cluster $[k]$ to $[1]$ or 1-local. The total number of substitutions per amino acid corresponds to $\sum_{k=2}^{D \log_2 N} \binom{D \log_2 N}{k} = N^{D}-D \log_2 N -1$.
To carry out the procedure for all $(N-2)$ amino acids the number of
ancilla qubits required is $(N-2)(N^{D}-D \log_2 N -1)$.
The number of qubits needed to represent a 2-local Hamiltonian version of the
protein Hamiltonian is given by adding the number of ancillary qubits to the number of original $(N-2) D \log_2 N$ quantum bits,
\begin{align}\label{eq:resources}
 \text{\# of total qubits for a 2-local expression} &= (N-2)(N^{D}-D \log_2 N -1) + (N-2) D \log_2 N \nonumber\\
& = (N-2)(N^D-1)
\end{align}
Eq.~\ref{eq:resources} provides a closed formula for the number of qubits needed to
find the lowest energy conformations for a protein with $N$ amino acids in $D$
dimensions in our encoding. In particular, for the case of a four amino acid peptide HPPH in
two dimensions considered in Sec.~\ref{sec:n4d2} requires 30 qubits.

\section{Conclusions}
We constructed the essential elements of an adiabatic quantum algorithm to
find the lowest energy conformations of a protein in a lattice model. The number
of binary variables needed to represent $N$ amino acids on an $N \times N$ lattice
is $(N-2) D \log_2 N$. The maximum locality of the final Hamiltonian, as determined by the interaction between pairs of amino acids using the mapping defined here, is $2 D \log_2 N$.

General strategies to construct energy functions to map into other quantum
mechanical Hamiltonians used for adiabatic quantum computing were presented. The
strategies used in the construction of the Hamiltonian for the HP model can be
used as general building blocks for Hamiltonians associated with physical systems where onsite
energies and/or pairwise potentials are present.

We also demonstrated an application of the Boolean scheme for converting a $k$-local Hamiltonian into a 2-local Hamiltonian, aiming toward an experimental implementation in quantum devices. The resulting
couplings, although 2-local, do not necessarily represent couplings among nearest
neighbor quantum bits in a two-dimensional geometry. It is however known that the number of ancillary physical qubits required to embed an arbitrary $N$ variable problem is upper-bounded by $N^2/(C-2)$, where $C$ is the number of couplers allowed per physical qubit.

The most important question remaining to be explored in future work is the scaling of run time $\tau$
with respect to the number of amino acids $N$. Run time $\tau$ is dependent on
the particular instance of the problem -- in our case, to different protein
sequences. It has been proposed that proteins have evolved towards a
many-dimensional funnel-like potential energy surface \cite{Creighton1992}. The
sequences that show a funnel-like structure might be easier to study using
adiabatic quantum computation, because the funnel structure may facilitate
annealing of the quantum wave function toward low energy conformations.

\begin{acknowledgments}
 We thank Jacob Biamonte, Sergio Boxio, Ivan Kassal, William Macready, Peter
 McMahon and  Rolando Somma for helpful discussions. Partial
 funding for this project was provided by a D-Wave Systems Inc. research
 contract and the Institute for Quantum Science and Enginnering at Harvard
 University.
\end{acknowledgments}

\appendix
\section{Extension of the pairwise interaction to  three dimensions and N
  amino acids , $N=2^M$ and $M\ge3$}

This extension follows the principles presented in
Sec.~\ref{subsubsec:pairwise} and extends the terms of the Hamiltonian to the
case of a three-dimensional lattice protein. The pairwise term for the
three-dimensional case is,
\begin{equation} \label{h-pairw-d3_ap}
H^{3D}_{pairwise}(N) = -\sum_{i,j=1}^{N} G_{ij}H^{ij,3D}_{pairwise}
\end{equation}

\begin{eqnarray}
x^{ij,3D}_{+}(N) = (1 - q_{f(i,1)+1})q_{f(j,1)+1} \prod_{s=2}^{\log_2 N}(1-q_{f(j,1)+s} \nonumber\\
-q_{f(i,1)+s} + 2\,q_{f(j,1)+s}\, q_{f(i,1)+s}) \prod_{s=1}^{\log_2 N} (1 - q_{f(i,2)+s} \nonumber\\
- q_{f(j,2)+s} + 2q_{f(i,2)+s}q_{f(j,2)+s}) \prod_{r=1}^{\log_2 N} (1 - q_{f(i,3)+r} \nonumber\\
- q_{f(j,3)+r} + 2q_{f(i,3)+r}q_{f(j,3)+r}),
\end{eqnarray}
\begin{eqnarray}
y^{ij,3D}_{+}(N) = (1 - q_{f(i,2)+1})q_{f(j,2)+1} \prod_{s=2}^{\log_2 N}(1-q_{f(j,2)+s} \nonumber\\
-q_{f(i,2)+s} + 2\,q_{f(j,2)+s}\, q_{f(i,2)+s}) \prod_{s=1}^{\log_2 N} (1 - q_{f(i,1)+s} \nonumber\\
- q_{f(j,1)+s} + 2q_{f(i,1)+s}q_{f(j,1)+s}) \prod_{r=1}^{\log_2 N} (1 - q_{f(i,3)+r} \nonumber\\
- q_{f(j,3)+r} + 2q_{f(i,3)+r}q_{f(j,3)+r}),
\end{eqnarray}
\begin{eqnarray}
z^{ij,3D}_{+}(N) = (1 - q_{f(i,3)+1})q_{f(j,3)+1} \prod_{s=2}^{\log_2 N}(1-q_{f(j,3)+s}\nonumber\\
-q_{f(i,3)+s} + 2\,q_{f(j,3)+s}\, q_{f(i,3)+s}) \prod_{s=1}^{\log_2 N} (1 - q_{f(i,1)+s} \nonumber\\
- q_{f(j,1)+s} + 2q_{f(i,1)+s}q_{f(j,1)+s}) \prod_{r=1}^{\log_2 N} (1 - q_{f(i,2)+r} \nonumber\\
- q_{f(j,2)+r} + 2q_{f(i,2)+r}q_{f(j,2)+r}),
\end{eqnarray}
\begin{eqnarray}
x^{ij,3D}_{-}(N) =  (1 - q_{f(i,1)+1})q_{f(j,1)+1} \Bigl[1-\prod_{k=1}^{\log_2 N}(1 - \nonumber\\
q_{f(i,1)+k})\Bigr] (q_{f(j,1)+2}+q_{f(i,1)+2} -  2\,q_{f(j,1)+2}\, q_{f(i,1)+2}) \nonumber\\
\prod_{r=3}^{\log_2 N} \Bigl[1 - (q_{f(j,1)+r} + \prod_{u=2}^{r-1} q_{f(j,1)+u} - 2 \prod_{u=2}^{r} q_{f(j,1)+u})\nonumber\\
 - q_{f(i,1)+r} + 2 q_{f(i,1)+r}(q_{f(j,1)+r} + \prod_{u=2}^{r-1} q_{f(j,1)+u} -  \nonumber\\
2 \prod_{u=2}^{r} q_{f(j,1)+u})\Bigr] \prod_{s=1}^{\log_2 N} (1 - q_{f(i,2)+s} - q_{f(j,2)+s} + \nonumber\\
2q_{f(i,2)+s}q_{f(j,2)+s}) \prod_{r=1}^{\log_2 N} (1 - q_{f(i,3)+r} - \nonumber\\
q_{f(j,3)+r} + 2q_{f(i,3)+r}q_{f(j,3)+r}),
\end{eqnarray}
\begin{eqnarray}
y^{ij,3D}_{-}(N) =  (1 - q_{f(i,2)+1})q_{f(j,2)+1} \Bigl[1-\prod_{k=1}^{\log_2 N}(1 - \nonumber\\
q_{f(i,2)+k})\Bigr] (q_{f(j,2)+2}+q_{f(i,2)+2} - 2\,q_{f(j,2)+2}\, q_{f(i,2)+2}) \nonumber\\
\prod_{r=3}^{\log_2 N} \Bigl[1 - (q_{f(j,2)+r} + \prod_{u=2}^{r-1} q_{f(j,2)+u} - \nonumber\\
2 \prod_{u=2}^{r} q_{f(j,2)+u}) - q_{f(i,2)+r} + 2 q_{f(i,2)+r} (q_{f(j,2)+r} + \nonumber\\
\prod_{u=2}^{r-1} q_{f(j,2)+u} - 2 \prod_{u=2}^{r} q_{f(j,2)+u})\Bigr] \nonumber\\
\prod_{s=1}^{\log_2 N} (1 - q_{f(i,1)+s} - q_{f(j,1)+s} + 2q_{f(i,1)+s}q_{f(j,1)+s})\nonumber\\
\prod_{r=1}^{\log_2 N} (1 - q_{f(i,3)+r} - q_{f(j,3)+r} + 2q_{f(i,3)+r}q_{f(j,3)+r}),
\end{eqnarray}
\begin{eqnarray}
z^{ij,3D}_{-}(N) =  (1 - q_{f(i,3)+1})q_{f(j,3)+1} \Bigl[1-\prod_{k=1}^{\log_2 N}(1 - \nonumber\\
q_{f(i,3)+k})\Bigr] (q_{f(j,3)+2}+q_{f(i,3)+2} - 2\,q_{f(j,3)+2}\, q_{f(i,3)+2}) \nonumber\\
\prod_{r=3}^{\log_2 N} \Bigl[1 - (q_{f(j,3)+r} + \prod_{u=2}^{r-1} q_{f(j,3)+u} - \nonumber\\
2 \prod_{u=2}^{r} q_{f(j,3)+u}) - q_{f(i,3)+r} + 2 q_{f(i,3)+r} (q_{f(j,3)+r} + \nonumber\\
\prod_{u=2}^{r-1} q_{f(j,3)+u} - 2 \prod_{u=2}^{r} q_{f(j,3)+u})\Bigr] \nonumber\\
\prod_{s=1}^{\log_2 N} (1 - q_{f(i,1)+s} - q_{f(j,1)+s} + 2q_{f(i,1)+s}q_{f(j,1)+s})\nonumber\\
\prod_{r=1}^{\log_2 N} (1 - q_{f(i,2)+r} - q_{f(j,2)+r} + 2q_{f(i,2)+r}q_{f(j,2)+r}).
\end{eqnarray}

\section{Truth table for the function resulting after the reduction of locality}\label{appendix:reduction}

\begingroup
\squeezetable
\begin{table}[h]
 \caption{\label{table:2localh} Truth table for the energy function $H_{toy,reduced}(\tilde q) =  1 + \tilde q_1 - \tilde q_2 + \tilde q_3 + \tilde q_4 -\tilde q_5 \tilde q_3 + \tilde q_5 \tilde q_6 +  \delta (3 \tilde q_5 + \tilde q_1 \tilde q_2 - 2 \tilde q_1 \tilde q_5 - 2 \tilde q_2 \tilde q_5) +\delta (3 \tilde q_6 + \tilde q_3 \tilde q_4 - 2 \tilde q_3 \tilde q_6 - 2 \tilde q_4 \tilde q_6)$. The top of the table shows the 16 non-penalized states that satisfy $\tilde q_5 = \tilde q_1 \wedge \tilde q_2$ and $\tilde q_6 = \tilde q_3 \wedge \tilde q_4$. These 16 states map one to one to the states in Table~\ref{table:4localh}. A sample of the remaning 48 penalized states are shown after the breaking line.}
\begin{ruledtabular}
\begin{tabular}{cccccc|c}
$\tilde q_6$ & $\tilde q_5$ & $\tilde q_4$ & $\tilde q_3$ & $\tilde q_2$ & $\tilde q_1$ & $H'(\tilde q_1,\tilde q_2,\tilde q_3,\tilde q_4,\tilde q_5,\tilde q_6)$. \\ \hline 0 & 0 & 0 & 0 & 1 & 0 & 0 \\ 0 & 0 & 0 & 0 & 0 & 0 & 1 \\ 0 & 1 & 0 & 0 & 1 & 1 & 1 \\ 0 & 0 & 0 & 1 & 1 & 0 & 1 \\ 0 & 0 & 0 & 1 & 1 & 1 & 1 \\ 0 & 0 & 1 & 0 & 1 & 0 & 1 \\ 0 & 0 & 0 & 0 & 0 & 1 & 2 \\ 0 & 0 & 0 & 1 & 0 & 0 & 2 \\ 0 & 0 & 1 & 0 & 0 & 0 & 2 \\ 0 & 1 & 1 & 0 & 1 & 1 & 2 \\ 1 & 0 & 1 & 1 & 1 & 0 & 2 \\ 0 & 0 & 0 & 1 & 0 & 1 & 3 \\ 0 & 0 & 1 & 0 & 0 & 1 & 3 \\ 1 & 0 &	 1 & 1 & 0 & 0 & 3 \\ 1 & 1 & 1 & 1 & 1 & 1 & 3 \\ 1 & 0 & 1 & 1 & 0 & 1 & 4 \\  \hline 0 &  1 &  0 &  0 &  1 &  0 &        $\delta$ \\ 0 &  1 &  0 &  1 &  1 &  0 &        $\delta$    \\
0 &  0 &  0 &  0 &  1 &  1 &        1 + $\delta$    \\
0 &  1 &  1 &  0 &  1 &  0 &        1 + $\delta$    \\
0 &  1 &  1 &  0 &  1 &  0 &        1 + $\delta$    \\
1 &  0 &  0 &  1 &  1 &  0 &        1 + $\delta$    \\
1 &  0 &  1 &  0 &  1 &  0 &        1 + $\delta$    \\
0 &  0 &  0 &  1 &  1 &  1 &        2 + $\delta$    \\
0 &  0 &  1 &  0 &  1 &  1 &        2 + $\delta$    \\
$\vdots$ &  $\vdots$ &  $\vdots$ &  $\vdots$ &  $\vdots$ &  $\vdots$ &        $\vdots$    \\
1 &  1 &  1 &  1 &  0 &  0 &        3 + 3 $\delta$    \\
1 &  0 &  0 &  0 &  1 &  1 &        1 + 4 $\delta$    \\
1 &  1 &  0 &  0 &  1 &  0 &        1 + 4 $\delta$    \\
1 &  1 &  0 &  1 &  0 &  0 &        2 + 4 $\delta$    \\
1 &  1 &  0 &  0 &  0 &  1 &        3 + 4 $\delta$    \\
1 &  1 &  1 &  0 &  0 &  0 &        3 + 4 $\delta$    \\
1 &  1 &  0 &  0 &  0 &  0 &        2 + 6 $\delta$    \\
\end{tabular}
\end{ruledtabular}
\end{table}
\endgroup


\end{document}